\documentclass[preprintnumbers,reprint,aps,twocolumn,showpacs,superscriptaddress,nofootinbib]{revtex4-2}
\usepackage{graphicx}
\usepackage{tabularx}
\usepackage{booktabs}
\usepackage{dcolumn}
\usepackage{bm}
\usepackage{amssymb}
\usepackage{amsmath}
\usepackage{multirow}
\usepackage{hyperref}
\usepackage{braket}
\usepackage{subcaption}
\usepackage{array}
\usepackage[math]{cellspace}
\usepackage[utf8]{inputenc}
\usepackage{physics}
\usepackage{cleveref}
\usepackage{url}
\hypersetup{
bookmarksnumbered,
colorlinks=true,
linkcolor=magenta,
urlcolor=magenta,
citecolor=blue
}
\DeclareCaptionJustification{blockjust}{%
  \leftskip=0pt \rightskip=0pt \parfillskip=0pt plus 1fil\relax}
\newcommand{\kaonp}{\raisebox{-0.4\totalheight}{\includegraphics[scale=.4]{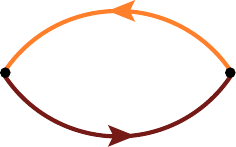}}}
\newcommand{\kaonn}{\raisebox{-0.4\totalheight}{\includegraphics[scale=.4]{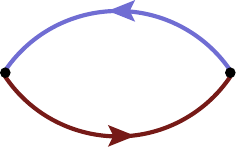}}}
\newcommand{\kaonlo}{\raisebox{-0.4\totalheight}{\includegraphics[scale=.4]{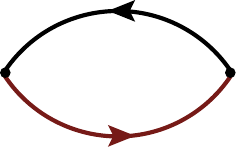}}}
\newcommand{\kaonnlo}{\raisebox{-0.4\totalheight}{\includegraphics[scale=.4]{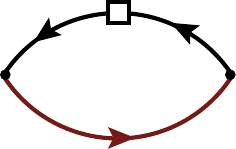}}}
\newcommand{\gisovlo}{\raisebox{-0.4\totalheight}{\includegraphics[scale=.4]{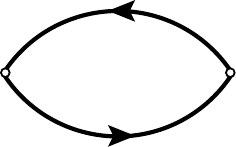}}}
\newcommand{\gisosconnss}{\raisebox{-0.4\totalheight}{\includegraphics[scale=.4]{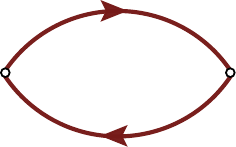}}}
\newcommand{\gisovfullu}{\raisebox{-0.4\totalheight}{\includegraphics[scale=.4]{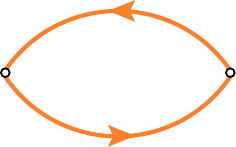}}}
\newcommand{\gisovfulld}{\raisebox{-0.4\totalheight}{\includegraphics[scale=.4]{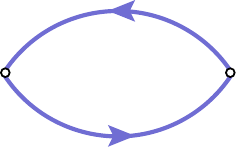}}}
\newcommand{\gibconn}{\raisebox{-0.4\totalheight}{\includegraphics[scale=.4]{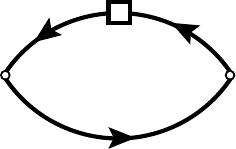}}}
\newcommand{\gibdiscll}{\raisebox{-0.1\totalheight}{\includegraphics[scale=.6]{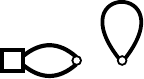}}}
\newcommand{\gibdiscls}{\raisebox{-0.1\totalheight}{\includegraphics[scale=.6]{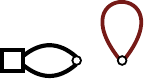}}}
\newcommand{\gibdiscuu}{\raisebox{-0.3\totalheight}{\includegraphics[scale=.5]{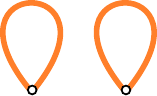}}}
\newcommand{\gibdiscdd}{\raisebox{-0.3\totalheight}{\includegraphics[scale=.5]{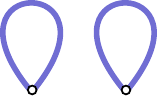}}}
\newcommand{\gibdiscus}{\raisebox{-0.3\totalheight}{\includegraphics[scale=.5]{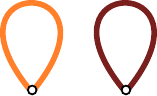}}}
\newcommand{\gibdiscds}{\raisebox{-0.3\totalheight}{\includegraphics[scale=.5]{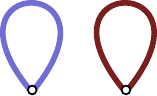}}}
\newcommand{\gibdisdiagS}{\raisebox{-0.3\totalheight}{\includegraphics[scale=.75,trim=0 0 27bp 20bp,clip]{figures/gibdiscll.pdf}}}
\newcommand{\gibdisdiag}{\raisebox{-0.3\totalheight}{\includegraphics[scale=.5]{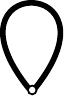}}}
\newcommand{\gisosdiscss}{\raisebox{-0.3\totalheight}{\includegraphics[scale=.5]{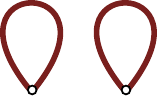}}}
\newcommand{\gisosdiscll}{\raisebox{-0.3\totalheight}{\includegraphics[scale=.5]{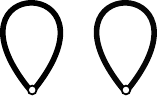}}}
\newcommand{\gisosdiscls}{\raisebox{-0.3\totalheight}{\includegraphics[scale=.5]{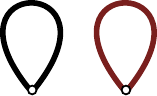}}}
\newcommand{\gisosdiscud}{\raisebox{-0.3\totalheight}{\includegraphics[scale=.5]{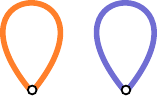}}}

\begin{document}
\preprint{MITP-26-068}
\preprint{CERN-TH-2026-290}

\title{Strong isospin breaking in the hadronic vacuum polarization:\\
       direct simulations versus Taylor expansion}

\date{\today}

\author{D.\ Albandea}
\email{albanded@uni-mainz.de}
\affiliation{Helmholtz-Institut Mainz, Staudingerweg 18, 55128 Mainz, Germany}
\affiliation{GSI Helmholtz Centre for Heavy Ion Research, 64291 Darmstadt, Germany}

\author{H.\ B.\ Meyer}
\affiliation{Helmholtz-Institut Mainz, Staudingerweg 18, 55128 Mainz, Germany}
\affiliation{Theoretical Physics Department, CERN, 1211 Geneva 23, Switzerland}
\affiliation{PRISMA$^{++}$ Cluster of Excellence and Institut f\"ur Kernphysik,
  Johannes Gutenberg-Universit\"at Mainz, 55099 Mainz, Germany}

\author{S.\ Kuberski}
\affiliation{Helmholtz-Institut Mainz, Staudingerweg 18, 55128 Mainz, Germany}
\affiliation{GSI Helmholtz Centre for Heavy Ion Research, 64291 Darmstadt, Germany}
\affiliation{PRISMA$^{++}$ Cluster of Excellence and Institut f\"ur Kernphysik,
Johannes Gutenberg-Universit\"at Mainz, 55099 Mainz, Germany}

\author{S.\ Lahrtz}
\affiliation{PRISMA$^{++}$ Cluster of Excellence and Institut f\"ur Kernphysik,
Johannes Gutenberg-Universit\"at Mainz, 55099 Mainz, Germany}

\author{H.\ Wittig}
\affiliation{Helmholtz-Institut Mainz, Staudingerweg 18, 55128 Mainz, Germany}
\affiliation{GSI Helmholtz Centre for Heavy Ion Research, 64291 Darmstadt, Germany}
\affiliation{PRISMA$^{++}$ Cluster of Excellence and Institut f\"ur Kernphysik,
Johannes Gutenberg-Universit\"at Mainz, 55099 Mainz, Germany}


\begin{abstract}
The inclusion of isospin-breaking effects is essential for high-precision
lattice QCD calculations, but the most efficient strategy to introduce them
remains an open methodological question. Focusing on strong isospin breaking, we
present a direct comparison of two approaches: an explicit simulation with
dynamical non-degenerate up and down quark masses ($N_f = 1+1+1$), and the
Taylor expansion in the up-down quark mass splitting $\Delta m$ on an
isospin-symmetric ($N_f = 2+1$) ensemble, using two CLS ensembles that differ
only in that splitting. We compute the kaon mass splitting and the isovector,
isoscalar and isospin-violating contributions to the hadronic vacuum
polarization (HVP) contribution to the anomalous magnetic moment of the muon,
together with $R_{38K}$, a renormalized, scheme-independent quantity that
normalizes the isospin-violating contribution by the kaon mass splitting. Even with
an up-down quark mass splitting twice as large as the physical one, we find that
the perturbative method at first order in $\Delta m$ agrees with the explicit
simulation for all observables considered, and that the dynamical treatment of
isospin breaking brings no gain in statistical precision at equivalent
statistics. This indicates that, at the current level of precision, combining
$N_f = 2+1$ ensembles with the perturbative expansion at first order in
$\Delta m$ is the computationally more efficient route to include strong
isospin-breaking corrections to the HVP.
\end{abstract}

\maketitle

\section{Introduction}

The anomalous magnetic moment of the muon, $a_\mu=(g-2)_\mu/2$,
provides one of the most precise tests of the Standard Model of
particle physics. The latest measurement campaign at Fermilab has
pushed the experimental precision to an impressive
124\,ppb~\cite{Muong-2:2006rrc,Muong-2:2021ojo,Muong-2:2023cdq,Muong-2:2025xyk},
which is likely to remain the benchmark for many years to come. At the
same time, the E34 experiment at J-PARC~\cite{Otani:2022wlj} aims at
an independent cross-check with completely different systematics. By
contrast, the uncertainty assigned to the Standard Model prediction as
summarized in the White Papers of 2020~\cite{Aoyama:2020ynm} and
2025~\cite{Aliberti:2025beg} is about 4\,times larger and is dominated
by hadronic effects. Among the latter, the hadronic vacuum
polarization (HVP) contribution plays a central role.

Lattice QCD offers a first-principles, systematically-improvable
method to compute the HVP contribution free of ambiguities arising
from experimentally measured hadronic cross sections that form the
basis of the traditional data-driven approach. The enormous and rapid
progress in the field is demonstrated by the fact that recent
calculations have reached sub-percent
precision~\cite{Borsanyi:2020mff,Djukanovic:2024cmq,Boccaletti:2024guq}.
At this level, it is mandatory to include the corrections due to the
isospin breaking arising from the mass difference between the up and
down quarks and from electromagnetic effects, which are known to
modify the HVP contribution at the percent level and are therefore essential in
order to make reliable Standard Model predictions.

As lattice determinations of the HVP aim at a precision of a few permille, developing
efficient strategies to incorporate isospin-breaking effects becomes an
important task.  For instance, such
corrections account for around half of the total uncertainty of the
LQCD prediction in Reference~\cite{Djukanovic:2024cmq}. Several methods have been
studied and used in the literature to include these effects. \emph{Exact
methods}, in which the isospin breaking contribution is included for both sea
and valence quarks, have been used to implement strong isospin breaking effects
through reweighting techniques~\cite{Finkenrath:2015ava} and to include
electromagnetic isospin breaking with dynamical QCD+QED
simulations~\cite{BMW:2013mpk,Boyle:2017gzv,Westin:2020blm,RCstar:2022yjz}.
\emph{Perturbative methods}, where isospin breaking is introduced as a Taylor
expansion in powers of the up-down quark mass splitting $\Delta m$ and the QED
coupling $e^2$ truncated to some power, were introduced in References~\cite{deDivitiis:2011eh,deDivitiis:2013xla}.
They have become the most commonly applied procedure to introduce
strong isospin breaking and electromagnetic effects in lattice QCD
simulations. We refer to this set of methods as RM123.

Including isospin-breaking corrections in lattice QCD simulations, however,
poses significant computational challenges and remains an active area of
research. Recently, Reference~\cite{RC:2025zoa} compared the efficiency of the
RM123 expansion for including QED with that of dynamical QCD+QED simulations
with $C^{*}$ boundary conditions, finding an error reduction by a factor of 3 in
the latter and that the main contribution to the uncertainty in RM123 comes from
disconnected sea-quark contributions. Alternative methods based on numerical
stochastic perturbation theory~\cite{DiRenzo:1994av,DallaBrida:2017tru}, such as
the one presented in Reference~\cite{Catumba:2023ulz}, achieve similar error
reductions by simulating a perturbative expansion of the theory with automatic
differentiation techniques.\footnote{Interestingly,
Reference~\cite{Catumba:2025ljd} suggests that a similar perturbative approach
can also be used to address the signal-to-noise problem.} Automatic
differentiation can also be applied directly by reweighting isospin-symmetric
ensembles, and a first study of strong isospin-breaking effects with this
technique has been explored in Reference~\cite{Albandea:2026cqh}.
Machine-learning techniques using normalizing flows to include higher-order
electromagnetic effects have been studied in
Reference~\cite{Hermansson-Truedsson:2026zlp}.

Similarly to the study performed in Reference~\cite{RC:2025zoa}, we aim to perform a
direct quantitative comparison between two complementary approaches to treat
strong isospin breaking effects at a common lattice spacing and quark masses:
the first consists in performing simulations with explicitly non-degenerate up
and down quark masses, while the second relies on performing a perturbative
expansion in the mass difference à la RM123~\cite{deDivitiis:2011eh}. The
comparison addresses two questions. First, since the explicit simulation
includes the mass splitting to all orders, it allows us to study the range of
validity of the perturbative expansion truncated at first order in $\Delta m$,
which is important in order to assess the robustness of the standard methods in
high-precision calculations. Second, by comparing observables that are sensitive
to strong isospin breaking, such as the kaon mass splitting and the HVP
contribution to $a_\mu$, we quantify the relative statistical efficiency of the
two approaches.

The paper is organized as follows. In Section~\ref{sec:rm123-method} we review the
RM123 method. In Section~\ref{sec:theory} we review the time-momentum
representation of the HVP and its decomposition into isospin components.
Section~\ref{sec:lattice} describes the lattice setup of the two ensembles used,
and the methods used to build the relevant correlators. Results are presented in
Section~\ref{sec:results}, followed by conclusions in Section~\ref{sec:conclusions}.

\section{The RM123 method}
\label{sec:rm123-method}

In lattice QCD simulations, we compute expectation values of a physical
observable $O$ by a statistical estimation of the path integral of
the theory. For the case of a world with 3 quarks, we are ideally interested in
evaluating the expectation value
\begin{align}
  \label{eq:intro:target-integral}
\langle O \rangle_{\Delta m} = \frac{\int \mathcal{D}U\mathcal{D}\psi\mathcal{D}\bar{\psi}\; O\;
e^{-S}}{\int
\mathcal{D}U\mathcal{D}\psi\mathcal{D}\bar{\psi} \; e^{-S}},
\end{align}
in a theory with $N_{f}=1+1+1$ non-degenerate
quarks $\psi =[u,d,s]$ corresponding to the up, down, and strange,
respectively, whose action reads
\begin{align}
  \label{eq:full-action}
  S = S_{G} + \sum_{f=u,d,s}^{}\sum_{x}^{} \bar{q}_{f} \left( D + m_{f} \right) q_{f}(x),
\end{align}
where $S_{G}$ is the action of the gauge fields $U$, $D$ is the massless Dirac
operator and $m_{f}$ are the bare quark masses. Introducing the average
light-quark mass $m_{l} = (m_{u}+m_{d})/2$ and the up-down mass
splitting $\Delta m = (m_{d}-m_{u})/2$, the light-quark mass term can be
rewritten as
\begin{align*}
  m_{u}\bar{u}u + m_{d}\bar{d}d = m_{l}\,[\bar{u}u + \bar{d}d] - \Delta m\,[\bar{u}u - \bar{d}d],
\end{align*}
so that the action splits into an isospin-symmetric and an isospin-breaking
part, $S = S_{0} + S_{\mathrm{IB}}(\Delta m)$, with
\begin{align}
                               \label{eq:sib-operator}
  S_0 =&\; S_{G} + \sum_{f=u,d,s}^{}\sum_{x}^{}\bar{q}_{f}D q_{f}(x) + m_{l} \sum_{x}^{} [\bar{u}u + \bar{d}d](x) \nonumber \\
  &+ m_{s}\sum_{x}^{}\bar{s}s(x), \\
  S_{\mathrm{IB}} =&\; -\Delta m \, O_{\mathrm{IB}},\,\,\,
  O_{\mathrm{IB}} \equiv \sum_{x}^{} [\bar{u}u - \bar{d}d](x). \nonumber
\end{align}
The whole dependence on the up-down mass splitting is therefore contained
in $S_{\mathrm{IB}}(\Delta m)$, which vanishes at $\Delta m = 0$.

While generating an ensemble with $N_{f}=2+1$ fermions (degenerate light
quarks) can be done efficiently by applying the Hybrid Monte Carlo (HMC)
algorithm~\cite{Duane:1987de} to the up/down quark doublet, generating an ensemble
with $N_{f}=1+1+1$ fermions (non-degenerate light quarks) is computationally more
expensive, as it requires the use of more complex algorithms such as the
Rational HMC (RHMC)~\cite{Kennedy:1998cu,Clark:2003na} for each of the light
quark flavors. An alternative method that allows one to circumvent the generation of
isospin-broken gauge ensembles, known as the RM123 method and introduced in
Reference~\cite{deDivitiis:2011eh}, consists of a perturbative expansion in powers
of $\Delta m$ of the path integral around the isosymmetric point $\Delta m = 0$.
Expanding Equation~(\ref{eq:intro:target-integral}) to first order in $\Delta m$, one
obtains
\begin{align}
  \label{eq:rm123:path-integral-expansion}
\langle O \rangle_{\Delta m} =&\; \frac{\int \mathcal{D}U\mathcal{D}\psi\mathcal{D}\bar{\psi}\; O\,(1+ \Delta m\, O_{\mathrm{IB}})
e^{-S_{0}}}{\int \mathcal{D}U\mathcal{D}\psi\mathcal{D}\bar{\psi}\; (1+ \Delta m\, O_{\mathrm{IB}}) e^{-S_{0}}} +
                                \mathcal{O}(\Delta m^2) \nonumber \\
  =&\; \frac{\langle O \rangle_{0} + \Delta
m\langle O\, O_{\mathrm{IB}} \rangle_{0}}{1 + \Delta m\langle O_{\mathrm{IB}}
\rangle_{0}} + \mathcal{O}(\Delta m^2) \nonumber \\
=&\; \langle O \rangle_{0} + \Delta m \langle O\, O_{\mathrm{IB}} \rangle_{0} + \mathcal{O}(\Delta m^2),
\end{align}
where we have used $e^{-S_{\mathrm{IB}}} = 1 + \Delta m\, O_{\mathrm{IB}} +
\mathcal{O}(\Delta m^{2})$ and $\langle O_{\mathrm{IB}} \rangle_{0} = 0$, since
in the isospin-symmetric theory there is perfect isospin symmetry. Note that the
expectation values on the right-hand side are evaluated in the theory
with $N_{f}=2+1$ with $\Delta m = 0$, which can be simulated more efficiently.
 This method is the standard
procedure to introduce strong (and electromagnetic) isospin breaking effects in
lattice QCD simulations.

However, we now need to evaluate new expectation values, like the
term $\langle O\, O_{\mathrm{IB}} \rangle_{0}$ in
Equation~(\ref{eq:rm123:path-integral-expansion}), which generate additional
diagrams. Their calculation requires additional inversions with scalar
insertions, including quark-disconnected contributions, which are generally
statistically noisier than the leading-order terms. Whether the overall balance
favors the perturbative expansion or a dedicated isospin-broken ensemble is
therefore not obvious a priori: the former avoids the cost of
generating $N_{f} = 1+1+1$ configurations, but pays for it at the level of the
measurements, while the latter includes the splitting to all orders
in $\Delta m$ at a higher cost per configuration.  Moreover, the truncation at
first order in Equation~(\ref{eq:rm123:path-integral-expansion}) introduces a
systematic uncertainty of $\mathcal{O}(\Delta m^{2})$ that cannot be assessed
within the first-order expansion. Quantifying both aspects for observables entering
the HVP contribution to $a_{\mu}$ is the purpose of this work.

\section{Hadronic vacuum polarization}
\label{sec:theory}

The HVP contribution to the anomalous magnetic moment of the muon can be
expressed as an integral over Euclidean momenta of the vacuum polarization
function $\Pi(Q^2)$ convoluted with a known kernel~\cite{DellaMorte:2017dyu},
\begin{align}
a_\mu^{\mathrm{hvp}} = 4\alpha^2 \int_0^\infty dQ^2\, K(Q^2; m_\mu^2)
\left\{ \Pi(Q^2) - \Pi(0) \right\},
\end{align}
where $\alpha$ is the fine-structure constant and $m_\mu$ the muon mass.
Using the time-momentum representation (TMR)~\cite{Bernecker:2011gh}, the
subtracted vacuum polarization is directly obtained from the spatially summed
two-point correlator
\begin{align}
\delta_{kl} G(t) = -\int d^3x\, \langle J_k^\gamma(t, \vec{x}) J_l^\gamma(0) \rangle,
\end{align}
where $J_{\mu}^{\gamma}$ is the electromagnetic current,
\begin{align}
J_{\mu}^{\gamma} = \frac{2}{3} \bar{u}\gamma_{\mu}u - \frac{1}{3}
\bar{d}\gamma_{\mu}d - \frac{1}{3} \bar{s}\gamma_{\mu}s + \dots
\end{align}
with the dots denoting the contribution from heavy quarks. With this, one can
obtain the compact expression~\cite{Djukanovic:2024cmq,Ce:2022kxy}
\begin{align}
a_\mu^{\mathrm{hvp}} = \left(\frac{\alpha}{\pi}\right)^2
\int_0^\infty dt\, \tilde{K}(t; m_\mu)\, G(t),
\label{eq:TMR}
\end{align}
where the analytically known kernel is
\begin{align}
\tilde{K}(t; m_\mu) = 4\pi^2 \int_0^\infty dQ^2\, K(Q^2; m_\mu^2)
\left[ t^2 - \frac{4}{Q^2} \sin^2\!\left(\frac{Qt}{2}\right) \right],
\end{align}
and
\begin{align}
K(Q^2; m_\mu^2) = \frac{m_\mu^{-2} \left(\sqrt{4m_\mu^2/Q^2+1}-1\right)^3
}{\left[4\left(\sqrt{4m_\mu^2/Q^2+1}+1\right)\sqrt{4m_\mu^2/Q^2+1}\right]}.
\end{align}

\subsection{Decomposition into flavor components and window observables}

It is convenient to decompose the electromagnetic current in terms of
the $\mathrm{SU}(3)$ flavor
currents $J_\mu^m = \bar\psi \frac{\lambda^m}{2} \gamma_\mu \psi$,
where $\lambda^m$ are the Gell--Mann matrices and the electromagnetic current
corresponds
to
$T^\gamma = \mathrm{diag}(\tfrac{2}{3}, -\tfrac{1}{3}, -\tfrac{1}{3}, \ldots)$.
For this study we will be particularly interested in the isovector and isoscalar
currents,
\begin{align}
  J^{3}_{\mu} =&\; \frac{1}{2}(\bar{u}\gamma_{\mu}u - \bar{d}\gamma_{\mu}d), \\
  J_{\mu}^{8} =&\; \frac{1}{2 \sqrt{3}}(\bar{u}\gamma_{\mu}u + \bar{d}\gamma_{\mu}d - 2 \bar{s}\gamma_{\mu}s).
\end{align}
For the renormalization and $\mathcal{O}(a)$ improvement of the correlators,
explained in Section~\ref{sec:renormalization}, we
will also need the flavor-singlet current, obtained with $\lambda^{0}$ the
identity,
\begin{align}
  J_{\mu}^{0} = \frac{1}{2}(\bar{u}\gamma_{\mu}u + \bar{d}\gamma_{\mu}d + \bar{s}\gamma_{\mu}s).
\end{align}
The full correlator $G^{(\gamma,\gamma)}$ decomposes as
\begin{align}
  \label{eq:gamma-gamma-decomposition}
G^{(\gamma,\gamma)} = G^{(3,3)} + \frac{1}{3} G^{(8,8)}
+ \frac{2}{\sqrt{3}} G^{(3,8)} + \dots,
\end{align}
and the generic correlator can be written as
\begin{align}
  \label{eq:generic-correlator-hvp}
\delta_{kl} G^{(m,n)}(t) = -\int d^3x\,
\langle J_k^m(t,\vec{x}) J_l^n(0) \rangle.
\end{align}
Separate TMR integrals can be defined for each component,
\begin{align}
a_\mu^{m,n} = \left(\frac{\alpha}{\pi}\right)^2
\int_0^\infty dt\, \tilde{K}(t; m_\mu)\, G^{(m,n)}(t),
\end{align}
where $a_\mu^{3,3}$ corresponds to the isovector contribution, $a_{\mu}^{8,8}$
to the isoscalar contribution, and $a_{\mu}^{3,8}$ to the isospin-violating
contribution to the HVP. In the following, we quote the isoscalar contribution in
the combination $\tfrac13 a_\mu^{8,8}$ in which it
enters $a_\mu^{\gamma,\gamma}$.

Furthermore, to enable clean comparisons between different lattice computations
and to isolate contributions from different distance scales, it is useful to
introduce windowed variants of the HVP~\cite{RBC:2018dos,Djukanovic:2024cmq}.
These are defined by inserting a smooth window function $W(t)$ into the TMR
integral,
\begin{align}
(a_\mu^{m,n})^X = \left(\frac{\alpha}{\pi}\right)^2
\int_0^\infty dt\, \tilde{K}(t; m_\mu)\, G^{(m,n)}(t)\, W^X(t),
\end{align}
where the windows are built from the smooth step function
$\Theta(t, d, \Delta) = \frac{1}{2}(1 + \tanh[(t-d)/\Delta])$ as
\begin{align}
W^{\mathrm{SD}} &= 1 - \Theta(t, d_0, \Delta), \nonumber \\
W^{\mathrm{ID}} &= \Theta(t, d_0, \Delta) - \Theta(t, d_1, \Delta), \\
W^{\mathrm{LD}} &= \Theta(t, d_1, \Delta), \nonumber
\end{align}
with standard parameters $\Delta = 0.15$~fm, $d_0 = 0.4$~fm, $d_1 =
1.0$~fm~\cite{RBC:2018dos} that isolate the short-distance (SD),
intermediate-distance (ID) and long-distance (LD) contributions.

Finally, in this work we also make use of a window that excludes the large-$t$
region where some correlators are no longer statistically resolved, in order to
maximize the significance of the higher-order isospin-breaking effects extracted
in Section~\ref{sec:lo-hvp}. For that purpose we define
\begin{align}
  \label{eq:def:ld-prime-window}
W^{\mathrm{LD}'} = \Theta(t, d_1, \Delta) - \Theta(t, d_2, \Delta),
\end{align}
with $d_2 = 1.5$~fm, which isolates the long-distance region between $1.0$
and $1.5$~fm. We denote it by LD$'$ and quote it, for completeness, alongside
the standard windows in our tables of results.

\subsection{Isospin-breaking contributions to the HVP and definition of $R_{38K}$}

In this work, we will use an ensemble with $N_{f}=2+1$ fermions, generated as
part of the CLS initiative~\cite{Bruno:2016plf,Bali:2016umi}, and therefore defined in
isoQCD.

It is useful to note that the isospin-breaking operator in
Equation~(\ref{eq:sib-operator}) is $G$-parity odd. Since $G^{(3,3)}$ and $G^{(8,8)}$
are $G$-parity even, their expansion only contains even powers of $\Delta m$,
and therefore their first strong isospin-breaking correction is of
order $\mathcal{O}(\Delta m^2)$. On the other hand, $G^{(3,8)}$ is $G$-parity
odd, so its expansion only contains odd powers of $\Delta m$ and is therefore
the only contribution with strong isospin breaking effects at
order $\mathcal{O}(\Delta m)$.

More generally, isospin-breaking corrections can be added perturbatively in the
parameters
\begin{align}
\varepsilon \equiv (\Delta m, e^2),
\end{align}
where $e$ is the electromagnetic coupling. Particularly, one can write the
isospin-violating contribution to the HVP as
\begin{align}
  \label{eq:a38-raw-expansion}
a_{\mu}^{3,8} = \frac{\partial a_{\mu}^{3,8}}{\partial e^2}e^2 + \frac{\partial a_{\mu}^{3,8}}{\partial \Delta m}\Delta m + \mathcal{O}(\varepsilon^2).
\end{align}
This quantity is among the largest isospin-breaking corrections to the HVP. In
QCD with $m_u \neq m_d$, it is the only flavor component of the HVP that
receives corrections at $\mathcal{O}(\Delta m)$. Once QED is included, however,
this separation is no longer unambiguous: the electromagnetic quark-mass
counterterms introduce scalar insertions in all flavor channels, and the value
of $\Delta m$ at the physical point depends on how these counterterms are fixed,
making comparisons between different collaborations difficult.

A meaningful comparison of the strong isospin-breaking contribution between
collaborations can nevertheless be built by considering the expansion of the
kaon mass splitting,
\begin{align}
  \Delta M_{K}^{\mathrm{phys}} = \frac{\partial \Delta M_{K}}{\partial e^2} e^2 + \frac{\partial \Delta M_{K}}{\partial \Delta m} \Delta m + \mathcal{O}(\varepsilon^2).
\end{align}
Isolating $\Delta m$ and inserting it into Equation~(\ref{eq:a38-raw-expansion}), one
obtains
\begin{align}
  \label{eq:amu38:renormalized}
  a_{\mu}^{3,8} =& \;   \frac{\partial a_{\mu}^{3,8}}{\partial e^2 } e^2 + R_{38K}  \cdot (\Delta M_{K}^{\text{phys}} - \Delta M_{K}^{\gamma}) + \mathcal{O}(\varepsilon^2),
\end{align}
where $\Delta M_{K}^{\gamma} \equiv e^2 \partial \Delta M_{K} / \partial e^2$
and we have defined the renormalized
quantity~\cite{Erb:2025nxk}
\begin{align}
  \label{eq:r38k-definition}
R_{38K} \equiv \frac{\partial a_{\mu}^{3,8}}{\partial \Delta M_{K}} = \frac{1}{\frac{\partial \Delta M_{K}}{\partial \Delta m}} \cdot \frac{\partial a_{\mu}^{3,8}}{\partial \Delta m} \Big|_{\mathrm{isoQCD}}.
\end{align}
This quantity is scheme-independent to first order in the
expansion $(e^2, \Delta m)$, and its determination and comparison between both the RM123
framework and an explicitly isospin-broken simulation is one of the main goals
of this work.\footnote{We note that a complete comparison between the two
  methods for the computation of $a_{\mu}^{3,8}$ in
  Equation~(\ref{eq:amu38:renormalized}) would require the computation of the first
  term, since the combination of electromagnetic and strong isospin breaking
  effects can lead to a significant cancellation of errors~\cite{RC:2025zoa}.
  Even though the inclusion of QED effects is beyond the scope of this work, we
  emphasize the importance of the computation of $R_{38K}$, as its scheme
independence makes a comparison between different collaborations possible.
}

Furthermore, FLAG defines pure QCD with $m_{u} \neq m_{d}$
at $\Delta M_{K} = -6$ MeV~\cite{FlavourLatticeAveragingGroupFLAG:2024oxs}.
Following this definition, we will quote the pure QCD contribution to the
isospin violating part of the HVP in this work as
\begin{align}
  \label{eq:flag-scheme}
a_{\mu,\mathrm{SIB}}^{3,8} \equiv - 6 \mathrm{\; MeV} \times R_{38K}.
\end{align}

\subsection{Spectral decomposition of $G^{(a,b)}$}

Generally, writing $\bar{J}_{\mu}^{a}(t)\equiv (a^3/L^{3/2})\sum_{\vec x} {J}_{\mu}^{a}(t,\vec x)$,
one can express the decomposition of $G^{(a,b)}(t; \Delta m)$ in finite volume as
\begin{align}
  G^{(a,b)}(t; \Delta m) =& -\bra{0}\bar{J}_{1}^{a}(t)\bar{J}_{1}^{b}(0)\ket{0}  \\
  =& -\sum_{n}^{}
 \bra{0}\bar{J}_{1}^{a}(0)\ket{n} \bra{n}\bar{J}_{1}^{b}(0)\ket{0} e^{-E_{n}t}, \nonumber
\end{align}
where we have summed over all unit-normalized states $I = \sum_{n}^{} \ket{n}\bra{n}$ and used
that $\bar J_{\mu}^{a}(t) = e^{Ht} \bar J_{\mu}^{a}(0)e^{-Ht}$.

The lowest-energy state coupling to the isovector current $\bar J_{1}^{3}$ ($I=1$)
in the ensembles of this study is close in energy to the $\rho$ meson\footnote{\label{fn:two-pion-threshold}On N451,
the $\rho$-like ground state lies below the lowest two-pion state in the
box: $am_\rho = 0.302(4)$ compared to $aE_{2\pi} = 0.3431(4)$, i.e.
about $792$~MeV and $900$~MeV respectively~\cite{Ce:2022eix}. The isovector
correlator is therefore saturated by a single state at the distances relevant
for our fits, in contrast to simulations at the physical pion mass in a 6\,fm box, where four
states are needed to describe the tail
at $t \gtrsim 1.5$~fm~\cite{Djukanovic:2024cmq}.}, while the lowest-energy state
coupling to the isoscalar $\bar J_{1}^{8}$ ($I=0$) is an $\omega$-meson-like state.
Thus we use the labels `$\rho$' and `$\omega$' to label the finite-volume states in the subsequent equations,
even if the relation of theses states to the corresponding mesons living in infinite volume is far from trivial.
At $\Delta m = 0$ there is perfect isospin symmetry, and the spectral
decompositions of the isovector and isoscalar correlators at leading order read
\begin{align}
G^{(3,3)}(t; \Delta m = 0) =& \, \left| \bra{0} \bar J_{1}^{3} \ket{\rho} \right|^2 e^{-E_{\rho}t} + \dots \\
  G^{(8,8)}(t; \Delta m = 0) =&\, \left| \bra{0} \bar J_{1}^{8} \ket{\omega} \right|^2 e^{-E_{\omega}t} + \dots \\
  G^{(3,8)}(t; \Delta m = 0) =&\; 0.
\end{align}
Since the $\rho$ and $\omega$ states are nearly, but not exactly degenerate,
the derivative of their respective energy levels with respect to $\Delta m$
vanishes at $\Delta m=0$.\footnote{This can be seen via the Feynman--Hellman
theorem, taking note that $O_{\mathrm{IB}}$ is a $G$-parity odd operator.}
Thus, to first order in $\Delta m$, we can treat their energies as unchanged from the isosymmetric point.
Both of them contribute at large distances,
\begin{align}
G^{(3,8)}(t; \Delta m) = A_{3\rho}^* A_{8\rho} e^{-E_{\rho}t} + A_{3\omega}^*
A_{8\omega} e^{-E_{\omega}t} + \dots
\end{align}
where $A_{a n} \equiv \bra{0}\bar{J}_{\mu}^{a}\ket{n}_{\Delta m}$. Since we are interested in
the first-order contribution in $\Delta m$, we take the derivative
\begin{align}
  \label{eq:g38-spec-dec-o1}
\partial_{\Delta m} G^{(3,8)}(t; \Delta m) \Big|_{\Delta m = 0} =&\; A_{3\rho}^* A_{8\rho}^{\prime}
  e^{-E_{\rho}t} \\
  & + A_{3\omega}^{*\,\prime} A_{8\omega} e^{-E_{\omega}t} \nonumber
\end{align}
and drop higher-lying states. Here and in the following, a prime denotes the
derivative $\partial_{\Delta m}$ evaluated at $\Delta m = 0$.
Due to the near-degeneracy of the $\rho$ and the $\omega$ states, a term
quasi-proportional to $t$  will also appear. To see this, we can obtain
$A_{a n}^{\prime}$
by expanding the eigenstate $\ket{n}_{\Delta m}$ using Rayleigh--Schrödinger
perturbation theory for states,
\begin{align}
\ket{n}_{\Delta m} = \ket{n} - \Delta m \sum_{k \neq n}^{} \ket{k}
\frac{\bra{k}O_{\mathrm{IB}}\ket{n}}{E_{n}-E_{k}},
\end{align}
where $O_{\mathrm{IB}}$ is the isospin-breaking operator defined in
Equation~(\ref{eq:sib-operator}). For the first term in the spectral
decomposition of Equation~(\ref{eq:g38-spec-dec-o1}), we will then have
\begin{align}
A_{3\rho}^* A_{8\rho}^{\prime} =& -A_{3\rho}^* \sum_{k \neq \rho}^{}
\frac{\bra{0}\bar{J}_{1}^{8}\ket{k}\,\bra{k}O_{\mathrm{IB}}\ket{\rho} }{E_{\rho}-
  E_{k}} \\
  =& A_{3\rho}^*A_{8\omega} \frac{s_{\omega\rho}}{E_{\rho}-E_{\omega}} +
a_{\rho},
\end{align}
where $s _{mn}\equiv -\bra{m}O_{\mathrm{IB}}\ket{n}$ and
\begin{align}
a_{\rho} \equiv A_{3\rho}^*\sum_{k\neq \rho,\omega}^{}A_{8k} \frac{s_{k\rho}}{E_{\rho}-E_{k}}.
\end{align}
Similarly, for the second term we have
\begin{align}
A_{3\omega}^{*\,\prime} A_{8\omega} = A_{3\rho}^* A_{8\omega} \frac{s
_{\rho\omega}^*}{E_{\omega}-E_{\rho}} + a_{\omega},
\end{align}
where
\begin{align}
a_{\omega} \equiv A_{8\omega}\sum_{k\neq \rho,\omega}^{}A_{3k}^* \frac{s _{k\omega}^*}{E_{\omega}-E_{k}}.
\end{align}
Combining both terms into the spectral decomposition of
Equation~(\ref{eq:g38-spec-dec-o1}), and using $s_{\rho\omega}^* = s_{\omega\rho}$ since $O_{\rm IB}$ is hermitian, we find
\begin{align}
\partial_{\Delta m}G^{(3,8)}(t; \Delta m)\Big|_{\Delta m = 0} =& \, A_{3\rho}^*
A_{8\omega} s _{\omega\rho} \frac{e^{-E_{\rho}t} -
  e^{-E_{\omega}t}}{E_{\rho}-E_{\omega}}  \nonumber \\
  &+ a_{\rho}e^{-E_{\rho}t} + a_{\omega}e^{-E_{\omega}t}.
\end{align}
Since the $\rho$ and $\omega$ states are nearly degenerate, we can
write $E_{\omega} = E_{\rho} + \Delta E$
and expand $e^{-E_{\omega}t} \approx e^{-E_{\rho}t}(1-\Delta E t)$. We finally find
\begin{align}
  \label{eq:g38-decomposition}
\partial_{\Delta m}G^{(3,8)}(t; \Delta m)\Big|_{\Delta m = 0} =  \left( A + Bt \right)e^{-E_{\rho}t},
\end{align}
where $A \equiv a_{\rho}+a_{\omega}$ and $B\equiv -A_{3\rho}^*A_{8\omega} s _{\omega\rho}$.
If for instance the gap $\Delta E$ is 10\,MeV, the approximation (\ref{eq:g38-decomposition}) breaks down at
$t\approx 20$\,fm, far beyond the reach of our lattice QCD data. It also means that until such a large time separation is reached,
the correlator (\ref{eq:g38-decomposition}) is not sensitive to the small energy scale $\Delta E$.

\section{Lattice setup}
\label{sec:lattice}

\subsection{Ensembles}
\label{sec:ensembles}

We use two CLS ensembles~\cite{Bruno:2016plf,Bali:2016umi} with $\mathcal{O}(a)$-improved
Wilson fermions and a tree-level improved L\"uscher-Weisz gauge
action~\cite{Bulava:2013cta}. The simulation parameters of both ensembles are
summarized in Table~\ref{tab:ensembles}. The two ensembles, labeled as N451 and
Y450, correspond to an $N_{f}=2+1$ simulation and an $N_{f}=1+1+1$ simulation,
respectively.

Both ensembles share the same lattice volume, bare gauge coupling and strange
hopping parameter, and differ only in the light quark sector. The up and down
hopping parameters of Y450 are tuned symmetrically about the N451 light quark mass,
$\tfrac{1}{2}(\kappa_u^{-1} + \kappa_d^{-1}) = \kappa_l^{-1}$, so that the trace
of the bare quark mass matrix, $\mathrm{tr}\;M = m_u + m_d + m_s$, is identical on
the two ensembles, and equal to its value along the CLS $N_f = 2+1$ trajectory
at this value of $\beta$~\cite{Bruno:2016plf}. This is relevant, on the one
hand, because the quark-mass dependence of the
$\mathcal{O}(a)$-improved bare
coupling, $\tilde{g}_0^2 = g_0^2(1 + b_g\,a \,\mathrm{tr}\;M)$, enters only
through $\mathrm{tr}\;M$, so both ensembles share the same improved coupling and
therefore the same lattice spacing (up to effects of $\mathcal{O}(\Delta m^2)$);
on the other hand, expanding around the isosymmetric point ensures the absence of
sea-quark disconnected diagrams to order $\Delta m$~\cite{deDivitiis:2011eh}.


\begin{table*}[t]
\begin{tabular}{lccccccccccc}
\toprule
Ensemble & $N_f$ & $\beta$ & Geometry & $\kappa_{l}$ & $\kappa_{s}$
   & $a\Delta m$ & $a$ [fm] & $M_{\pi}$ [MeV] & $\bar{M}_{K}$ [MeV]
  & $\Delta M_{K}$ [MeV]  & $N_{\mathrm{cfg}}$  \\
\midrule
N451 & $2+1$ & $3.46$ & $48^3 \times 128$ & $0.1370616$ & $0.1365480771$
     & $0$ & 0.075 & 291 & 468 & 0 & 1011  \\
Y450 & $1+1+1$ & $3.46$ & $48^3 \times 128$ & $0.1370616$ & $0.1365480771$
     & $0.00068596$ & 0.075 & 291 & 468 & 14 & 450  \\
\bottomrule
\end{tabular}
\caption{CLS ensembles used in this study. $N_f$ denotes the number of
dynamical quark flavors, $N_{\mathrm{cfg}}$ the number of gauge
configurations, and $a$ the lattice spacing determined from the gradient
flow scale $t_0$. Both ensembles share the same isosymmetric light and
strange hopping parameters. In Y450 the up and down quarks are split
symmetrically about the light point,
$1/(2\kappa_{u,d}) = 1/(2\kappa_{l}) \mp a\Delta m$, corresponding to
$\kappa_{u} = 0.1370873775536536$ and $\kappa_{d} = 0.137035832138812$,
which leaves $\mathrm{tr}\; M$ unchanged.}
\label{tab:ensembles}
\end{table*}

The two ensembles are simulated with different algorithms.
In N451 the degenerate light quarks form a doublet whose
determinant $\det(D^{\dagger}D)$ is simulated efficiently with an even-odd
preconditioned twisted-mass Hasenbusch frequency
splitting~\cite{Hasenbusch:2001ne,Luscher:2008tw}, with the lightest factor
regulated by a twisted mass $\mu_0$ that is subsequently removed by reweighting;
the single strange quark is treated with the RHMC
algorithm~\cite{Kennedy:1998cu,Clark:2003na}. In Y450 the up and down quarks are
non-degenerate, so each of the three flavors is simulated with
a separate RHMC factorization (for details of the simulation, see
Appendix~\ref{app:rhmc}). Because the light-quark Dirac operator reaches smaller
eigenvalues than the strange one, its rational approximation requires a higher
degree and a lower cutoff than in the strange sector. Consequently, going from
the $N_f = 2+1$ to the $N_f = 1+1+1$ setup replaces one comparatively cheap
twisted-mass doublet by two separate single-flavor RHMC factorizations, with
roughly $2.5$ times the number of pseudofermion fields, which is the dominant
overhead of the explicit approach at the level of ensemble
generation.\footnote{We emphasize, nonetheless, that the cost of generating a
gauge configuration is not by itself a figure of merit, since an ensemble is
generated once and reused for many observables. At the level of the computation
of correlation functions in this work, both approaches require the same number
of inversions of the Dirac operator and therefore share a similar computational
cost.} In practice, N451
was generated with version~1.6 of the \texttt{openQCD}
code~\cite{Luscher:2012av,Luscher2019_openQCD}, while Y450 was generated with
version~2.4.2.

The amount of ``physical'' isospin breaking can be seen from the computation of the
kaon mass splitting. In the case of the isospin-broken ensemble Y450, this
computation is trivially done to all orders in $\Delta m$ by computing the
correlation function of the charged and neutral kaons,
\begin{align}
  \label{eq:explicit:kaon-corr}
  C_{K^{+}K^{-}} =&\;  \kaonp = \mathrm{tr} \left[ \gamma_{5} D^{-1}(m_{u}) \gamma_{5}D^{-1}(m_{s}) \right], \\
  C_{K^{0}K^{0}} =&\; \kaonn = \mathrm{tr} \left[ \gamma_{5} D^{-1}(m_{d}) \gamma_{5}D^{-1}(m_{s}) \right], \nonumber
\end{align}
where the filled dot represents a $\gamma_{5}$ insertion and $D^{-1}(m_{f})$ is
the inverse Wilson Dirac operator of the quark flavor $f$ ($u$ orange, $d$
purple, $s$ red).

Analogously, in the isospin-symmetric N451 ensemble one can compute the kaon
mass splitting with the RM123 approach to first order in $\Delta m$,
\begin{align}
  \label{eq:rm123:kaon-corr-expansion}
  C_{K^{+}K^{-}} =  \; \kaonlo + \kaonnlo \;\Delta m + \mathcal{O}(\Delta m^2), \\
  C_{K^{0}K^{0}} =  \; \kaonlo - \kaonnlo \;\Delta m + \mathcal{O}(\Delta m^2),
\end{align}
where the black lines represent $D^{-1}(m_{l})$, the empty square represents a
scalar current insertion\footnote{We note that diagrams with a scalar current
insertion can be alternatively computed by finite differences, e.g.\ by
evaluating the correlators of Eq.~(\ref{eq:explicit:kaon-corr}) on the
isospin-symmetric ensemble with valence quark masses
$m_{u,d} = m_l \mp \Delta m$, and dividing their difference by $2\Delta m$, as
done in Reference~\cite{Borsanyi:2020mff} for the strong isospin-breaking
corrections to hadron masses and to the quark-disconnected contribution to the
HVP. Both approaches coincide in the limit $\Delta m \to 0$, and are also
equivalent to computing the derivative exactly with automatic
differentiation~\cite{Albandea:2026cqh}.} coming from
$S_{\mathrm{IB}}$, and
\begin{align}
  \kaonlo =& \;\mathrm{tr} \left[ \gamma_{5} D^{-1}(m_{l}) \gamma_{5}D^{-1}(m_{s}) \right], \label{eq:meson-2pt} \\ \kaonnlo =& \; \mathrm{tr} \left[ \gamma_{5} D^{-1}(m_{l}) D^{-1}(m_{l}) \gamma_{5} D^{-1}(m_s) \right]. \label{eq:rm123:o1}
\end{align}

\begin{figure}[t]
    \centering
    \includegraphics[width=0.49\textwidth]{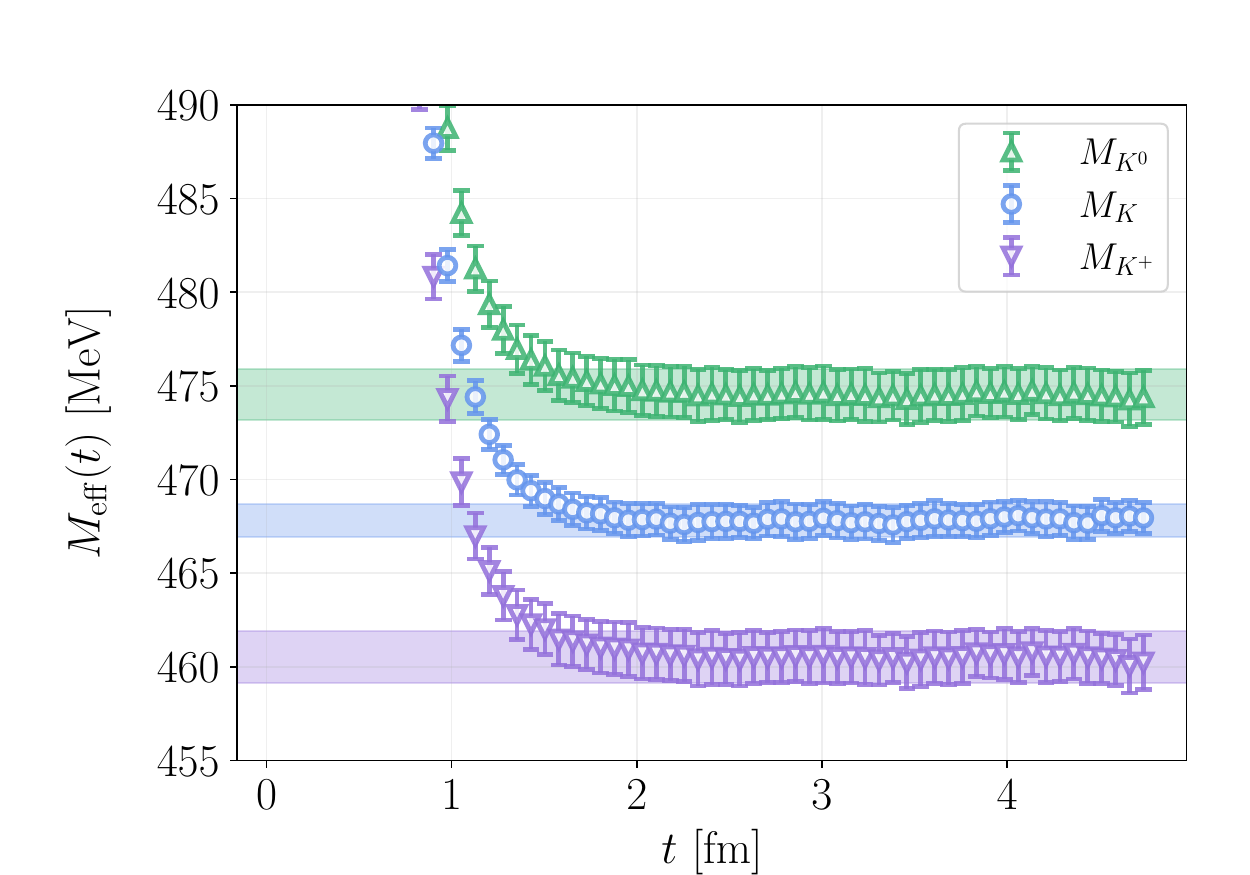}
    \caption{Effective mass of the charged kaon (purple, downward triangles),
      neutral kaon (green, upward triangles) as computed in the $N_{f}=1+1+1$
      ensemble, and the isoQCD kaon (blue circles) from the $N_{f}=2+1$
      ensemble. Bands represent plateau averages.}
    \label{fig:kaon-effective-masses}
\end{figure}

In Figure~\ref{fig:kaon-effective-masses} we show the effective masses for the
charged and neutral kaons in the $N_{f}=1+1+1$ ensemble and the isoQCD kaon in
the $N_{f}=2+1$, computed from the correlators above and using the standard
functional form for the effective mass with periodic boundary conditions,
\begin{align}
aM_{\mathrm{eff}}(t) = \cosh^{-1} \frac{C(t + a) + C(t-a)}{2 C(t)}.
\end{align}
The measured kaon mass splitting at our
chosen $\Delta m$ is
\begin{align}
  \label{eq:kaon-splitting-result-y450}
M_{K^{\pm}} - M_{K^{0}}  = &\; -14.026(52) \mathrm{\; MeV} \quad \text{ (Y450)}.
\end{align}
Comparing to the scheme for pure, isospin-broken QCD defined in the FLAG
review~\cite{FlavourLatticeAveragingGroupFLAG:2024oxs},
\begin{align}
  \label{eq:kaon-splitting-flag}
 M_{K^{\pm}} - M_{K^{0}} \equiv&\; - 6 \mathrm{\; MeV} \quad \text{ (FLAG)},
\end{align}
the ensemble Y450 is more than twice the physical one.  However, we note
that this bigger splitting was intended: rather than reproducing the true
splitting, our focus is to study whether higher order strong isospin breaking
effects can be detected at the current level of precision, for which a bigger
splitting can be advantageous.

\subsection{Correlator construction}
\label{sec:correlator-construction}

We will focus on the computation of the isovector ($G^{(3,3)}$), isoscalar
($G^{(8,8)}$), and the isospin-violating ($G^{(3,8)}$) contributions to the HVP.

In practice, we average over the three spatial directions,
\begin{align}
  \label{eq:Gk-average}
  G^{(m,n)} \equiv \frac{1}{3} \sum_{k=1}^{3} G_{k}^{(m,n)},
\end{align}
where $G_{k}^{(m,n)}$ is the $k=l$ component of
Eq.~(\ref{eq:generic-correlator-hvp}). The isovector correlator receives no
$\mathcal{O}(\Delta m)$ correction, so to first order in $\Delta m$ it can be
computed as\footnote{Our convention is not to include the minus sign associated with a fermion loop in the diagram.}
\begin{align}
  \label{eq:g33-rm123}
G^{(3,3)}_{k} = \frac{2}{4} \times \gisovlo + \mathcal{O}(\Delta m^2),
\end{align}
where the empty dot denotes an insertion of $\gamma_{k}$.
Alternatively, in the
isospin broken ensemble it can be computed (to all orders in $\Delta m$) as
\begin{align}
    \label{eq:g33-disconnected-y450}
  G^{(3,3)}_{k} =&\; \frac{1}{4} \times \left[  \gisovfullu + \gisovfulld \right. \\
  &\left. - \gibdiscuu - \gibdiscdd + 2\times \gisosdiscud \right], \nonumber
\end{align}
where the disconnected contribution enters only at $\mathcal{O}(\Delta m^2)$ in
the isosymmetric theory.

Similarly, the isoscalar correlator to first order in $\Delta m$ reads
\begin{align}
  \label{eq:g88-diagrams}
  G^{(8,8)}_{k} =&\; \frac{1}{12} \times \left\{  2 \times \gisovlo + 4\times \gisosconnss \right. \\
                 & \left. - 4\times \left[  \gisosdiscll + \gisosdiscss - 2 \times \gisosdiscls \right] \right\}  \nonumber \\
 &+ \mathcal{O}(\Delta m^2), \nonumber
\end{align}
while in the non-degenerate theory it can be computed with
\begin{align}
  G^{(8,8)}_{k} =&\; \frac{1}{12} \times \left\{   \gisovfullu + \gisovfulld + 4\times \gisosconnss \right. \nonumber \\
  &  - \gibdiscuu - \gibdiscdd - 4\times \gisosdiscss \\
                 & \left. - 2\times \gisosdiscud + 4\times \left[ \gibdiscus + \gibdiscds \right] \right\}. \nonumber
\end{align}

Finally, the isospin-violating contribution to the HVP reads
\begin{align}
  G_{k}^{(3,8)} &=\; \frac{4}{4 \sqrt{3}} \times \left[  \gibconn  \right. \\
  & \left. -\; \gibdiscll +  \gibdiscls \right] \Delta m +\mathcal{O}(\Delta m^2), \nonumber
\end{align}
while for explicit isospin broken simulations
\begin{align}
  G_{k}^{(3,8)} =&  \; \frac{1}{4\sqrt{3}} \times \left[   \gisovfullu - \gisovfulld - \gibdiscuu   \right. \nonumber \\
  & \left. +\gibdiscdd + 2\times \gibdiscus -2\times \gibdiscds \right].
\end{align}

The renormalization and $\mathcal{O}(a)$ improvement of these correlators,
discussed in Section~\ref{sec:renormalization}, additionally require
correlators involving the flavor-singlet current $J_{\mu}^{0}$. In
particular, the
mixed singlet-octet correlator reads, to first order in $\Delta m$,
\begin{align}
  G^{(0,8)}_{k} =&\; \frac{2}{4\sqrt{3}} \times \left[ \gisovlo - \gisosconnss - 2\times \gisosdiscll \right. \nonumber \\
                 & \left. + \gisosdiscss + \gisosdiscls \right] + \mathcal{O}(\Delta m^2),
\end{align}
while to all orders in $\Delta m$ it can be computed as
\begin{align}
  G^{(0,8)}_{k} =&\; \frac{1}{4\sqrt{3}} \times \left[ \gisovfullu + \gisovfulld - 2\times \gisosconnss \right. \nonumber \\
                 & - \gibdiscuu - \gibdiscdd + 2\times \gisosdiscss \\
                 & \left. - 2\times \gisosdiscud + \gibdiscus + \gibdiscds \right]. \nonumber
\end{align}
Similarly, the mixed isovector-singlet correlator, which as $G^{(3,8)}$ starts
at $\mathcal{O}(\Delta m)$, reads, to first order in $\Delta m$,
\begin{align}
  G_{k}^{(3,0)} =&\; \frac{2}{4} \times \left[ 2\times \gibconn \right. \\
  & \left. - 2\times \gibdiscll - \gibdiscls \right] \Delta m +\mathcal{O}(\Delta m^2), \nonumber
\end{align}
while to all orders in $\Delta m$
\begin{align}
  G_{k}^{(3,0)} =&  \; \frac{1}{4} \times \left[   \gisovfullu - \gisovfulld - \gibdiscuu   \right. \nonumber \\
  & \left. +\gibdiscdd - \gibdiscus + \gibdiscds \right].
\end{align}

For the two point connected diagrams, the truncated solver
method~\cite{Bali:2009hu} is employed. This method is based on the identity
\begin{align}
  \label{eq:truncated-solver-method}
  \langle O^{\mathrm{exact}} \rangle =&\; \frac{1}{N_{\mathrm{sloppy}}} \sum_{i=1}^{N_{\mathrm{sloppy}}} \langle O_{i}^{\mathrm{sloppy}} \rangle \\
  &+ \frac{1}{N_{\mathrm{bc}}} \sum_{i=1}^{N_{\mathrm{bc}}} \left[ \langle O_{i}^{\mathrm{exact}} - O_{i}^{\mathrm{sloppy}} \rangle \right], \nonumber
\end{align}
where $\langle \dots \rangle$ denotes ensemble average. In our
case, $O_{i}^{\mathrm{sloppy}}$ is a stochastic determination of the trace in,
for example, Equation~(\ref{eq:meson-2pt}), where the Dirac operator is inverted to
low precision with $N_{\mathrm{sloppy}} = 256$ spin- and color-diluted stochastic time-slice sources
with U(1) noise. The low-precision solves
correspond to 9 iterations of the deflated and preconditioned solver~\cite{Luscher:2007es,Luscher:2007se,Luscher:2003qa} as implemented in the
\texttt{openQCD} code~\cite{Luscher2019_openQCD}. Lastly, the second term in
Equation~(\ref{eq:truncated-solver-method}) is a bias-correction term,
where $O^{\mathrm{exact}}_{i}$ corresponds to the computation of the same
observable with a high-precision inversion of the Dirac operator,
with $N_{\mathrm{bc}} = 8$.

For the disconnected quark loops we need to evaluate, for each flavor $f$, the
zero-momentum projected single-propagator traces
\begin{align}
  \label{eq:disc-loop}
\gibdisdiag =  \sum_{\vec{x}}^{} \mathrm{tr}
  \left[ \gamma_{k} D^{-1}(m_{f})(x,x) \right].
\end{align}
We employ a variant of the
frequency-splitting method of Reference~\cite{Giusti:2019kff}, as implemented in
Reference~\cite{Ce:2022eix}, which combines the one-end trick
(OET)~\cite{McNeile:2006bz} with the generalized hopping parameter expansion
(gHPE)~\cite{Gulpers:2013uca} and hierarchical probing
(HP)~\cite{Stathopoulos:2013aci}.

Following Reference~\cite{Ce:2022eix}, we use the chain of flavor of increasing mass $l \to s \to i \to c$ on
N451, where $i$ denotes an intermediate valence quark with
$\kappa_{i} = 0.133590$, and $\kappa_{c} = 0.125439$. On Y450 the
degenerate light quark is replaced by the two non-degenerate flavors $u$ and
$d$, resulting in the longer chain $u \to d \to s \to i \to c$ and doubling the
number of light-quark inversions.  We use $N_{s} = 512$ U(1) stochastic volume
sources for the $l-s$ hop on N451 and double this number for each subsequent,
heavier hop, i.e. $N_{s} = 1024$ for $s-i$ and $N_{s} = 2048$ for $i-c$. On Y450
the additional $u-d$ hop is evaluated with the same statistics
as $d-s$, $N_{s} = 512$.\footnote{We note that the $u-d$ difference is more strongly
correlated than $l-s$, so it could plausibly be estimated with fewer sources,
but we have not attempted to optimize this.}

For the disconnected quark loops with a scalar insertion required by the RM123
expansion,
\begin{align}
  \label{eq:disc-loop-scalar}
\gibdisdiagS =  \sum_{\vec{x}, z} \mathrm{tr}
  \left[ \gamma_{k} D^{-1}(m_{l})(x,z) D^{-1}(m_{l})(z,x) \right],
\end{align}
we use $N_{\mathrm{sloppy}}=128$ and $N_{\mathrm{bc}} = 8$
four-dimensional stochastic volume sources per configuration, placed at the
mass-insertion vertex. Using $\gamma_{5}$-hermiticity we construct a split-even
estimator, along the lines of Reference~\cite{Giusti:2019kff}.

\subsection{Renormalization, $\mathcal{O}(a)$ improvement and scale setting}
\label{sec:renormalization}

The renormalization of the observables considered in this work, including
mass-dependent $\mathcal{O}(a)$ effects and with three non-degenerate quark
masses, can be derived from the renormalization pattern of quark bilinears given
in Ref.~\cite{Bhattacharya:2005rb},
\begin{align}
  \label{eq:ren-pattern-nonsinglet}
\mathrm{tr}(\lambda V_{\mu})_{R} =& Z_{V} \left[
(1 + a\bar{b}_{V}  \tr M)\,\mathrm{tr}(\lambda V_{\mu})^{I}
  \right. \\
  &+ \left. \tfrac{1}{2}ab_{V}\,\mathrm{tr}(\{\lambda, M\}V_{\mu})
+ af_{V}\,\mathrm{tr}(\lambda M)\,\mathrm{tr}V_{\mu} \right], \nonumber
\end{align}
with
\begin{align}
  \tr(\lambda V_{\mu})^{I} =& \tr(\lambda V_{\mu}) + ac_{V} \tilde{\partial}_{\nu} \tr(\lambda T_{\mu\nu}), \\
  \sigma_{\mu\nu} =& -\frac{1}{2}[\gamma_{\mu},\gamma_{\nu}],
\end{align}
where we consider the local lattice discretization of the vector
current, $(V_{\mu})_{ij} = \bar{\psi}_{j}\gamma_{\mu}\psi_{i}$, and of the
tensor operator, $(T_{\mu\nu})_{ij} = \bar{\psi}_{j}\sigma_{\mu\nu}\psi_{i}$,
$\lambda$ is a traceless $3\times 3$ flavor matrix,
$M = \mathrm{diag}(m_{u}, m_{d}, m_{s})$ are the subtracted bare masses
\begin{align}
m_{i} = m_{0,i} - m_{\mathrm{cr}}, \quad i = u, d, s,
\end{align}
and we consider the symmetric lattice derivative
\begin{align}
\tilde{\partial}_{\nu}\phi(x) = \frac{\phi(x + a\hat{\nu}) - \phi(x-a\hat{\nu})}{2a}.
\end{align}
Here, $Z_{V}$, $b_{V}$, $\bar{b}_{V}$ and $f_{V}$ are functions of the $\mathcal{O}(a)$-improved
bare coupling only,
\begin{align}
\tilde{g}_{0}^{2} = g_{0}^{2}(1+ab_{g} \tr M).
\end{align}

We denote by $G_{I}^{(m,n)}$ the vector correlators defined as
in Eq.~(\ref{eq:generic-correlator-hvp}) but with the replacement
$J^{m}_{\mu} \to J^{m,I}_{\mu} \equiv \frac{1}{2}\tr(\lambda^{m}V_{\mu})^{I}$.
At zero spatial momentum, only the $\nu = 0$
component of the tensor term contributes and
\begin{align}
  G_{I}^{(m,n)}(t) =&\; G^{(m,n)}(t) + ac_{V}\tilde{\partial}_{t}\left[G_{TV}^{(m,n)}(t) + G_{TV}^{(n,m)}(t)\right] \nonumber \\
  &+ \mathcal{O}(a^2),
\end{align}
where we have used time-reflection symmetry to place the tensor at the sink,
with the tensor-vector correlator
\begin{align}
\delta_{kl} G_{TV}^{(m,n)}(t) = -\int d^3x\,
\langle T_{k0}^{m}(t,\vec{x}) J_l^n(0) \rangle.
\end{align}
where $T_{\mu\nu}^{m} = \bar{\psi}\frac{\lambda^{m}}{2}\sigma_{\mu\nu}\psi$. In practice,
$G^{(m,n)}$ are the raw correlators computed along the lines of
Sec.~\ref{sec:correlator-construction}, while $G_{TV}^{(m,n)}$ is
computed from the same connected and disconnected diagrams as $G_{k}^{(m,n)}$, with
the same coefficients, but with $\gamma_{k}$ replaced by
$\sigma_{k0} = \gamma_{0}\gamma_{k}$ at the sink.

Considering the renormalization pattern in
Eq.~(\ref{eq:ren-pattern-nonsinglet})---and neglecting
$f_{V}$~\cite{Bhattacharya:2005rb}, which is of $\mathcal{O}(g_0^6)$---, the renormalized and
$\mathcal{O}(a)$-improved correlators that we consider in this work read, to all
orders in $\Delta m$,
\begin{align}
G^{(3,3)}_{R} =&\; Z_{3}^{2}\,G_{I}^{(3,3)}
- Z_{V}^{2}\,b_{V}a\Delta m\left[\frac{2}{\sqrt{3}}G_{I}^{(3,8)} + \frac{4}{3}G_{I}^{(3,0)}\right]
                 \nonumber \\
  &+ \mathcal{O}(a^{2}), \label{eq:GR-33}\\
G^{(8,8)}_{R} =&\; Z_{8}^{2}\,G_{I}^{(8,8)} \label{eq:GR-88}
                 + Z_{V}^{2}\,b_{V}a\left[\frac{4(m_{l}-m_{s})}{3\sqrt{3}}G_{I}^{(0,8)} \right. \nonumber \\
                 &- \left. \frac{2\Delta m}{\sqrt{3}}G_{I}^{(3,8)}\right]
+ \mathcal{O}(a^{2}), \\
G^{(3,8)}_{R} =&\; Z_{3}Z_{8}\,G_{I}^{(3,8)}
+ Z_{V}^{2}\,b_{V}a\bigg[\frac{2(m_{l}-m_{s})}{3\sqrt{3}}G_{I}^{(3,0)} \nonumber \\
               &- \frac{\Delta m}{\sqrt{3}}\left(G_{I}^{(3,3)} + G_{I}^{(8,8)} + \frac{2}{\sqrt{3}}G_{I}^{(0,8)}\right)\bigg] \label{eq:GR-38}\\
  &+ \mathcal{O}(a^{2}), \nonumber
\end{align}
with
\begin{align}
Z_{3} =&\; Z_{V}\left[1 + \bar{b}_{V}a(2m_{l}+m_{s}) + b_{V}am_{l}\right], \\
Z_{8} =&\; Z_{V}\left[1 + \bar{b}_{V}a(2m_{l}+m_{s}) + \tfrac{1}{3}b_{V}a(m_{l}+2m_{s})\right]. \nonumber
\end{align}
Note that, since $2 m_{l} = m_{u} + m_{d}$, the same renormalization factor is
applied for both $N_{f}=2+1$ and $N_{f}=1+1+1$ ensembles used in this work.
Particularly, we use the renormalization constant $Z_V = 0.71998(6)$ and
the $\mathcal{O}(a)$-improvement
coefficients
${\binoppenalty=10000 \relpenalty=10000 b_V = 1.622(14)}$,
${\binoppenalty=10000 \relpenalty=10000 \bar{b}_V = 0.125(5)}$~\cite{Gerardin:2018kpy},
and $c_{V} = -0.180(13)$, obtained by evaluating the interpolation in
$g_0^2$ of Ref.~\cite{Harris:2025xvk} at $\beta = 3.46$, with the
uncertainty propagated from the covariance of its parameters.\footnote{The
  interpolation reads
  \begin{align*}
    c_{V}(g_{0}^{2}) = -0.01030\, C_{F}\, g_{0}^{2}
    \left[1 + (a + b\, g_{0}^{2})\, e^{-1/(2 b_{0} g_{0}^{2})}\right],
  \end{align*}
  with $C_{F} = 4/3$, $b_{0} = 9/(4\pi)^{2}$, $a = -2514.80$,
  $b = 2044.53$, and $g_{0}^{2} = 6/\beta$.}
The subtracted bare masses entering $Z_{3}$, $Z_{8}$ and the
mixing terms are $am_{i} = \frac{1}{2\kappa_{i}} - \frac{1}{2\kappa_{\mathrm{cr}}}$,
with $\kappa_{\mathrm{cr}} = 0.1370657(28)$ at $\beta = 3.46$~\cite{RQCD:2022xux}.
The lattice spacing is set via the gradient flow scale $t_0$, with
$t_0/a^2 = 3.682(7)$~\cite{Kuberski:2024bcj} and
$\sqrt{8t_0} = 0.4081(42)$~fm~\cite{Strassberger:2021tsu}, giving
$a \approx 0.0752$~fm.  Statistical errors are estimated using the $\Gamma$
method~\cite{Madras:1988ei,Wolff:2003sm} as implemented in
\texttt{ADerrors.jl}~\cite{Ramos:2018vgu}.

Finally, since the leading contributions of $G^{(3,0)}$ and
and $G^{(3,8)}$ start at $\mathcal{O}(\Delta m)$, some of the terms in
Equations~(\ref{eq:GR-33})--(\ref{eq:GR-38}) are of $\mathcal{O}(\Delta m^2)$.
To stay coherent with the first-order expansion in $\Delta m$ of the RM123
method we use instead, for the isoQCD ensemble,
\begin{align}
G^{(3,3)}_{R} =&\; Z_{3}^{2}\,G_{I}^{(3,3)} + \mathcal{O}(a^{2}, \Delta m^{2}), \\
G^{(8,8)}_{R} =&\; Z_{8}^{2}\,G_{I}^{(8,8)}
+ Z_{V}^{2}\,b_{V}a\,\frac{4(m_{l}-m_{s})}{3\sqrt{3}}G_{I}^{(0,8)} \nonumber \\
               &+ \mathcal{O}(a^{2}, \Delta m^{2}), \\
G^{(3,8)}_{R} =&\; Z_{3}Z_{8}\,G_{I}^{(3,8)}
+ Z_{V}^{2}\,b_{V}a\bigg[\frac{2(m_{l}-m_{s})}{3\sqrt{3}}G_{I}^{(3,0)} \nonumber \\
               &- \frac{\Delta m}{\sqrt{3}}\left(G_{I}^{(3,3)} + G_{I}^{(8,8)} + \frac{2}{\sqrt{3}}G_{I}^{(0,8)}\right)\bigg]  \\
               &+ \mathcal{O}(a^{2}, \Delta m^{2}).\nonumber
\end{align}

\section{Results}
\label{sec:results}

\subsection{Kaon mass splitting}

As briefly mentioned in Section~\ref{sec:ensembles}, the most direct observable to
compare the two methods of this study is the kaon mass splitting. Particularly,
it is interesting to check whether the explicit simulation with $N_{f}=1+1+1$
can lead to a more precise estimation of the
derivative $\partial \Delta M_{K} / \partial \Delta m$.

\begin{figure*}[t]
  \begin{subfigure}[t]{0.49\textwidth}
    \centering
    \includegraphics[width=\textwidth]{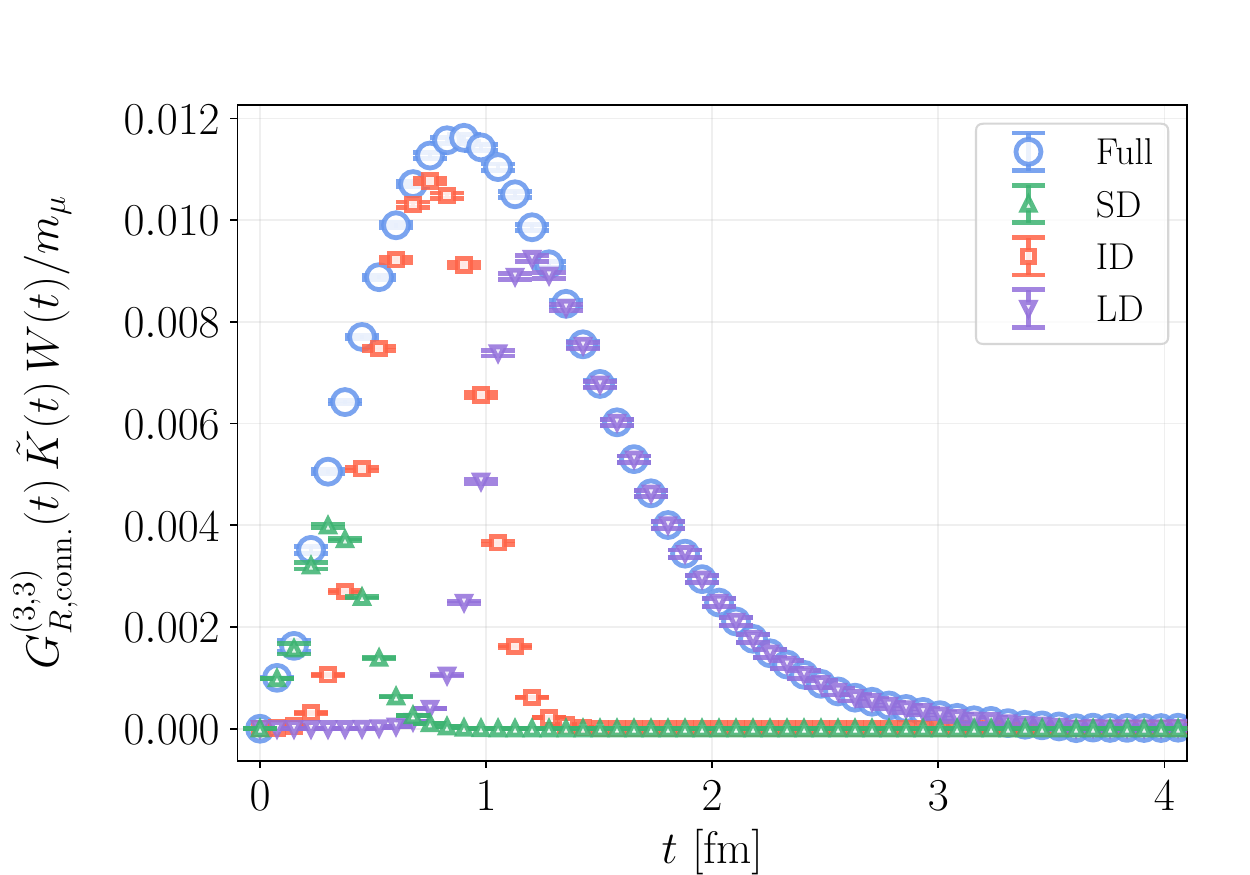}
  \end{subfigure}
  \hfill
  \begin{subfigure}[t]{0.49\textwidth}
    \centering
    \includegraphics[width=\textwidth]{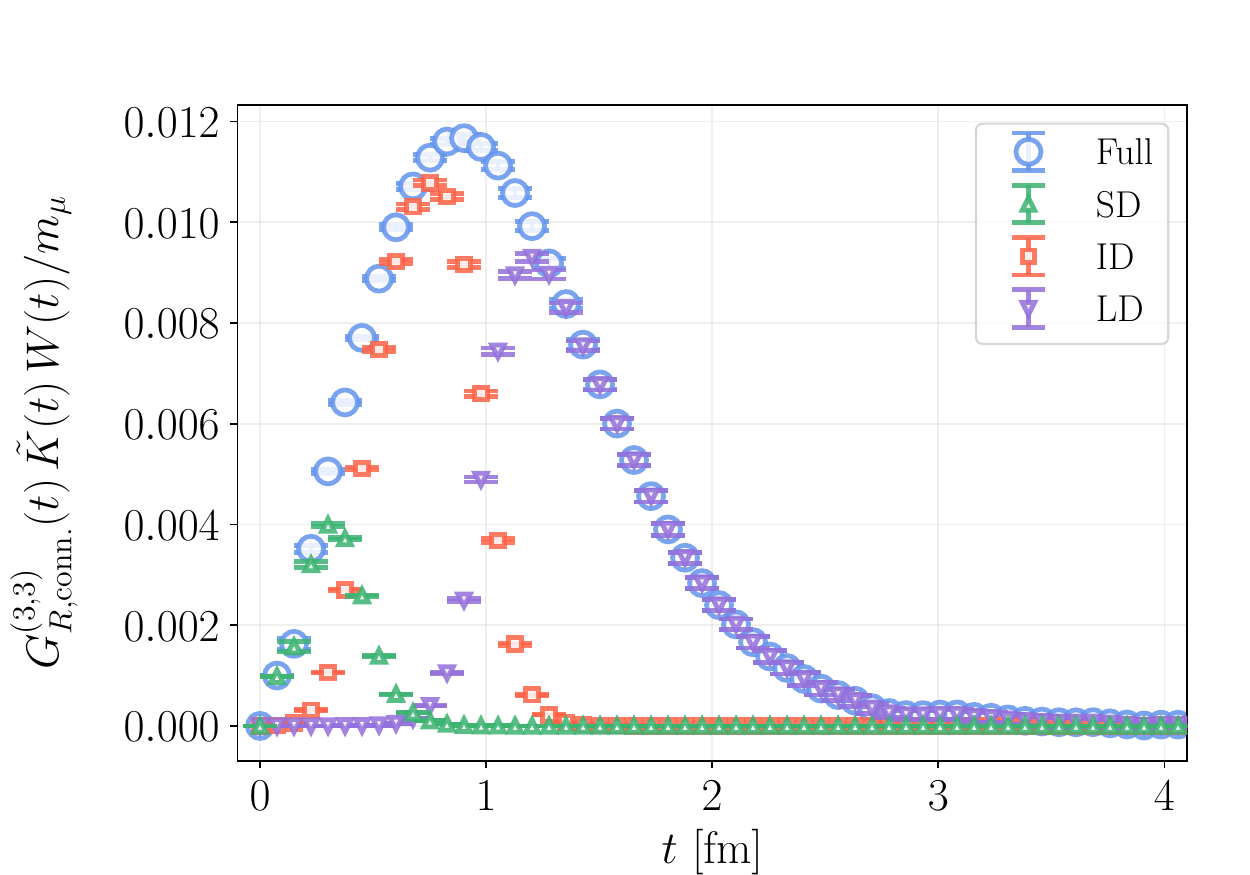}
  \end{subfigure}
  \caption{Integrand $G_{R,\mathrm{conn.}}^{(3,3)}(t)\,\tilde{K}(t)\,W(t)/m_{\mu}$,
    made dimensionless by the muon mass, as a function of Euclidean time $t$
    for the N451 (left) and Y450 (right) ensembles. The four sets of points correspond
    to the full (blue circles), short-distance (green upward triangles),
    intermediate-distance (red squares) and long-distance (purple downward
triangles) windows.}
  \label{fig:integrand-G33}
\end{figure*}

On the one hand, in isoQCD with periodic boundary conditions, one can extract
the leading order mass by fitting to
\begin{align}
C(t) = A \cosh \left( M_{K}(t - T / 2) \right),
\end{align}
while the first order in $\Delta m$ can be fit to the functional form
\begin{align}
  C^{\prime}(t) =&\; A^{\prime} \cosh \left( M_{K}(t - T / 2) \right)  \\
  &+ A\, M_{K}^{\prime}\,(t - T / 2) \sinh \left( M_{K}(t - T / 2) \right), \nonumber
\end{align}
where $C(t)$ and $C^{\prime}(t)$ are obtained in the RM123 method from
Equation~(\ref{eq:meson-2pt}) and Equation~(\ref{eq:rm123:o1}), respectively. With this,
we obtain a value for the derivative
$M_{K}^{\prime} = \partial M_{K^{\pm}}/\partial \Delta m
  = \partial \Delta M_{K}/\partial(2\Delta m)$ at $\Delta m = 0$
\begin{align}
  \label{eq:rm123-mK-result}
M_{K}^{\prime} = -3.9061(90).
\end{align}

On the other hand, in the explicit simulation one can compute the mass of the
charged kaon and neutral kaon to all orders with the correlators in
Equation~(\ref{eq:explicit:kaon-corr}). We can approximate the derivative by
finite differences to obtain
\begin{align}
\frac{\partial \Delta M_K}{\partial (2 \Delta m)}\Big|_{\Delta m = 0} \approx \frac{M_{K^{\pm}} - M_{K^0}}{2 \Delta m} =  -3.8960(140).
\end{align}
Comparing with Equation~(\ref{eq:rm123-mK-result}), we see  that there is good
agreement between the two methods, indicating that higher order effects
in $\Delta m$ are not resolved at the current level of statistics.
The two uncertainties, however, are not directly comparable, since N451 and Y450
differ in the number of configurations ($N_{\mathrm{cfg}} = 1011$ and $450$,
respectively). Rescaling the RM123 error to the statistics of Y450, i.e.
multiplying it by $\sqrt{1011/450} \approx 1.5$, brings the two determinations
to the same level of precision. The explicit simulation therefore shows no gain
in statistical precision for this observable.

Finally, note that one could also approximate the kaon splitting derivative by
taking finite differences between $M_{K}$ computed in the $N_{f}=2+1$ ensemble
and $M_{K^{0}}$ computed in the $N_{f}=1+1+1$ ensemble. This way, one would obtain
\begin{align}
\frac{\partial \Delta M_K}{\partial (2\Delta m)}\Big|_{\Delta m = 0} \approx \frac{M_{K}|_{\Delta m = 0} - M_{K^0}|_{\Delta m}}{\Delta m} =  -3.74(83),
\end{align}
which has a much bigger error than the derivatives computed within the same
ensemble and thus indicates the importance of exploiting correlations.

\begin{figure*}[t]
  \begin{subfigure}[t]{0.49\textwidth}
    \centering
    \includegraphics[width=\textwidth]{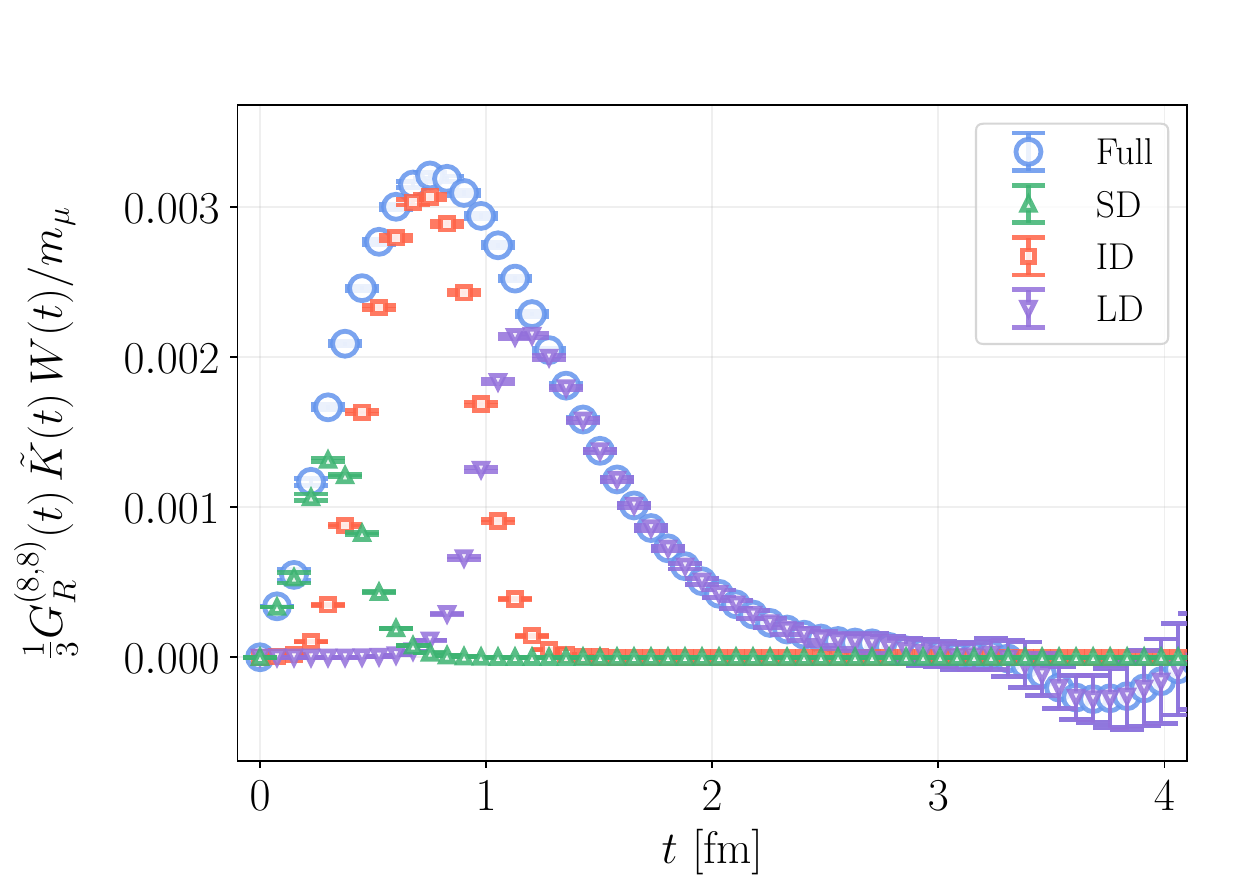}
  \end{subfigure}
  \hfill
  \begin{subfigure}[t]{0.49\textwidth}
    \centering
    \includegraphics[width=\textwidth]{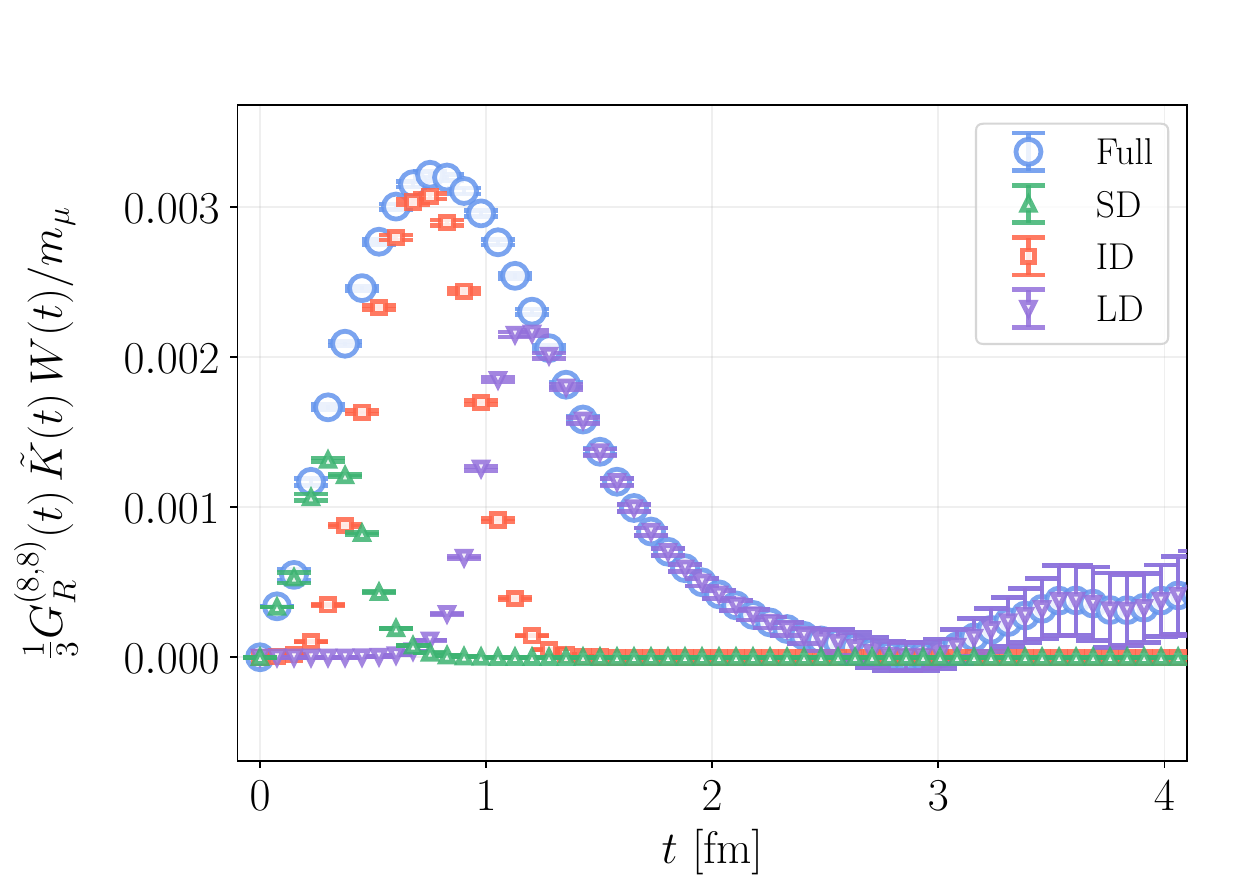}
  \end{subfigure}
  \caption{Integrand $\frac{1}{3} G_{R}^{(8,8)}(t)\,\tilde{K}(t)\,W(t)/m_{\mu}$
    as a function of Euclidean time $t$
    for the N451 (left) and Y450 (right) ensembles. For the explanation of
    symbols, see Figure~\ref{fig:integrand-G33}.}
  \label{fig:integrand-G88}
\end{figure*}

\begin{table*}[t]
\centering
\begin{tabular}{llrrrrr}
\toprule
 & & & \multicolumn{4}{c}{Window} \\
\cmidrule(lr){4-7}
 & Ens. & Full & SD & ID & LD & LD$'$ \\
\midrule
\multirow{2}{*}{$a_{\mu,\mathrm{conn.}}^{3,3}$} & N451 & $449.4(4.4)$ & $40.20(53)$   & $172.47(89)$  & $236.7(3.6)$ & $132.75(86)$ \\
                               & Y450 & $443.7(6.0)$ & $40.21(53)$   & $172.84(94)$  & $230.7(5.5)$ & $133.3(1.3)$ \\
\addlinespace
$a_{\mu,\mathrm{disc.}}^{3,3}$ & Y450 & $-1.8(3.4)$  & $0.00168(19)$ & $-0.0108(29)$ & $-1.8(3.4)$  & $-0.075(12)$ \\
\addlinespace
$a_{\mu}^{3,3}$                & Y450 & $441.9(6.5)$ & $40.21(53)$   & $172.83(94)$  & $228.9(6.0)$ & $133.3(1.3)$ \\
\addlinespace
\multirow{2}{*}{$\frac{1}{3} a_{\mu}^{8,8}$} & N451 & $113.7(6.6)$ & $13.10(19)$   & $49.41(30)$   & $51.2(6.5)$  & $30.80(22)$           \\
                               & Y450 & $101.5(10.0)$ & $13.10(19)$  & $49.51(31)$   & $38.9(9.9)$  & $30.94(26)$           \\
\bottomrule
\end{tabular}
\caption{HVP contributions $a_\mu^{3,3}$ (connected)  and $\frac{1}{3} a_{\mu}^{8,8}$, in units of
$10^{-10}$, for the full integral and its window decomposition, plus the
window ``LD$'$'' defined in Equation~(\ref{eq:def:ld-prime-window}),
computed with the renormalized and $\mathcal{O}(a)$-improved correlators $G_{R}$.
The $a_{\mu,\mathrm{disc.}}^{3,3}$ row gives the disconnected contribution
to $a_\mu^{3,3}$, which vanishes identically in isoQCD and is therefore quoted
for Y450 only.}
\label{tab:results:hvp-33-38}
\end{table*}

\subsection{Leading order contribution to the HVP}
\label{sec:lo-hvp}

The leading order isovector and isoscalar components account for
most of the HVP contribution to $g-2$. As outlined in
Section~\ref{sec:correlator-construction}, these do not have any contribution at
order $\mathcal{O}(\Delta m)$ and therefore provide a good test for higher-order
strong isospin-breaking effects.

Figure~\ref{fig:integrand-G33} shows the TMR integrand
$G_{R,\mathrm{conn.}}^{(3,3)}(t)\tilde{K}(t)W(t) / m_{\mu}$, corresponding to
the connected diagrams in Equations~(\ref{eq:g33-rm123})
and~(\ref{eq:g33-disconnected-y450}), as a function of Euclidean time $t$ for
the full integral and all three windows on both ensembles. In each panel the
four sets of points correspond to the full, SD, ID and LD windows, illustrating
how each window suppresses different time regions. The SD integrand peaks at
short distances ($t \lesssim 0.4$~fm), where the data is more accurate but
cutoff effects are most sizeable; the LD integrand is dominated by the tail at
$t \gtrsim 1.0$~fm where statistical errors are more prominent; and the ID
window captures the intermediate region where the balance between cutoff effects
and the signal-to-noise ratio is most favorable.

\begin{table*}[t]
\centering
\begin{tabular}{llrrrrr}
\toprule
 & & & \multicolumn{4}{c}{Window} \\
\cmidrule(lr){4-7}
 & Ens. & Full & SD & ID & LD & LD$'$ \\
\midrule
\multirow{2}{*}{$a_{\mu,\mathrm{conn.}}^{3,3}$} & N451 & $444.0(3.3)$ & $40.20(53)$   & $172.47(89)$ & $231.8(2.3)$ & $132.71(86)$ \\
                               & Y450 & $443.0(4.4)$ & $40.21(53)$   & $172.84(94)$ & $230.4(3.7)$ & $133.2(1.2)$  \\
\addlinespace
\multirow{2}{*}{$\frac{1}{3} a_{\mu}^{8,8}$} & N451 & $111.2(1.1)$ & $13.10(19)$   & $49.41(30)$  & $48.71(89)$  & $30.80(22)$   \\
                               & Y450 & $111.0(1.1)$ & $13.11(19)$   & $49.51(31)$  & $48.41(90)$  & $30.94(26)$   \\
\bottomrule
\end{tabular}
\caption{HVP contributions $a_\mu^{3,3}$ (connected)  and $\frac{1}{3} a_{\mu}^{8,8}$, in units
of $10^{-10}$, obtained with the bounding method~\cite{RBC:2018dos,Ce:2022eix}
from the renormalized and $\mathcal{O}(a)$-improved correlators $G_{R}$.}
\label{tab:results:hvp-bounding}
\end{table*}

Analogously, Figure~\ref{fig:integrand-G88} shows the
integrand $\frac{1}{3} G_{R}^{(8,8)}(t)\tilde{K}(t)W(t) / m_{\mu}$ as a function
of Euclidean time $t$ for all three windows on both ensembles. As we saw in
Equation~(\ref{eq:g88-diagrams}), this correlator includes quark-disconnected loops
that have a worse signal-to-noise problem at large $t$, and we can see that
indeed the signal degradation at large $t$ is more substantial than the one
in $G_{R}^{(3,3)}$.

Table~\ref{tab:results:hvp-33-38} collects the TMR integrals for the isovector contribution,
$a_\mu^{3,3}$, and isoscalar contribution, $\frac{1}{3} a_{\mu}^{8,8}$ to the HVP, computed
for both ensembles and for the three windows. Again, we see good agreement
between both ensembles, both having similar errors when accounting
for equivalent statistics.\footnote{We note that the short-distance results have
  comparable errors on the two ensembles, although N451 has roughly twice the
  statistics of Y450, which could lead one to conclude that the $N_f = 1+1+1$
  ensemble resolves the short-distance physics better. These errors, however,
scale as $\sqrt{2\tau_{\rm int}/N_{\rm cfg}}$, and for these observables the
integrated autocorrelation times lie close to the uncorrelated limit on both
ensembles and are determined with large relative uncertainties. The smaller
statistics of Y450 is partly compensated by a smaller autocorrelation factor,
even though both autocorrelation times are compatible with each other. Therefore, no
conclusion can be drawn from this coincidence.\label{fn:autocorr-barely-resolved}} Furthermore,
these results indicate that the sum of
the $\mathcal{O}(\Delta m^2)$ and higher-order contributions is no larger than
the uncertainty on the difference between the two ensembles. Assuming that most
of these higher-order contributions come from $\mathcal{O}(\Delta M_{K}^{2})$ effects, we can rescale
such uncertainty from the simulated kaon
mass splitting in Equation~(\ref{eq:kaon-splitting-result-y450}) to the FLAG
convention in Equation~(\ref{eq:kaon-splitting-flag}) by a
factor $(6/14.026)^2$, obtaining
\begin{align}
  \label{eq:a33-bound-dm2}
  a^{(3,3)}_{\mu, \mathcal{O}(\Delta M_{K}^2)} \lesssim 1 \times 10^{-10}.
\end{align}


\begin{figure}[t]
    \centering
    \includegraphics[width=0.49\textwidth]{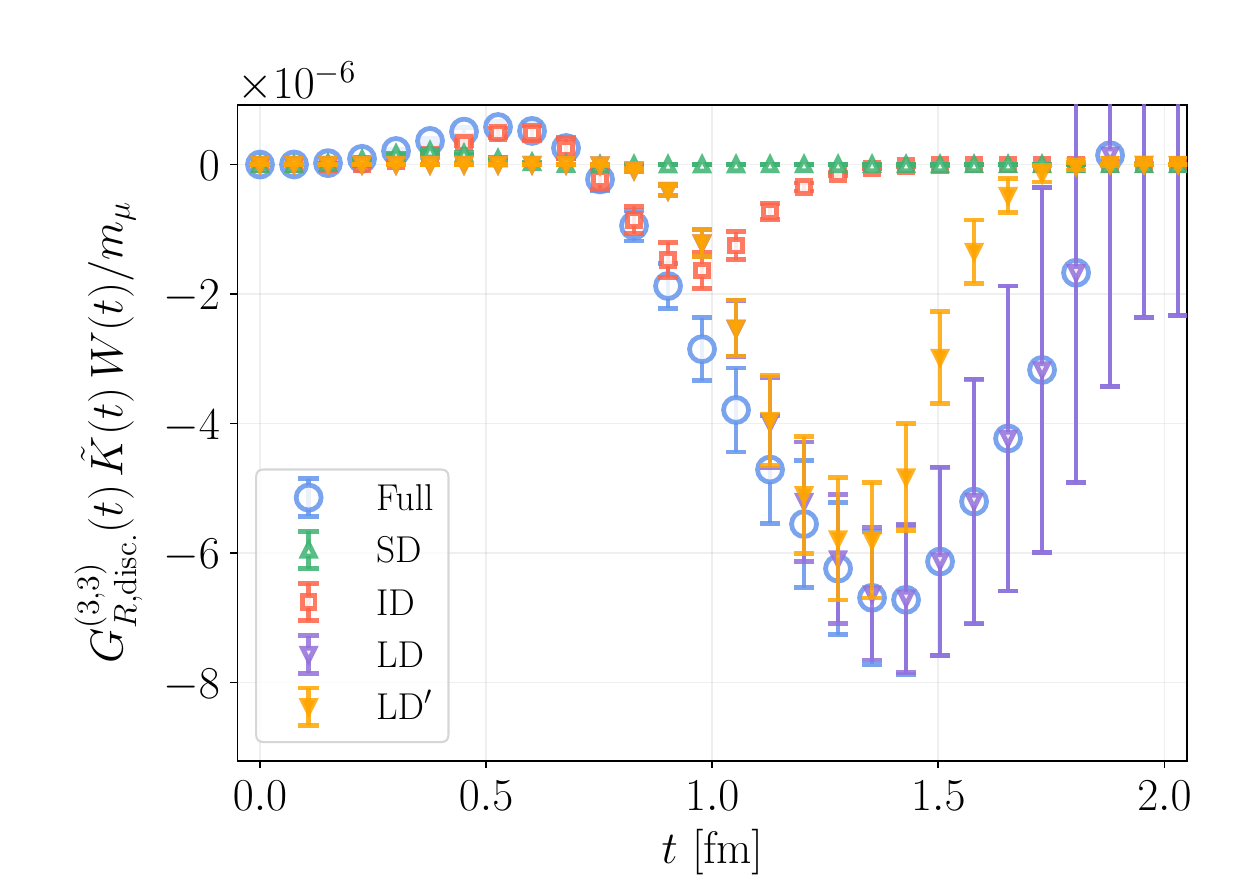}
  \caption{Integrand $G_{R,\mathrm{disc.}}^{(3,3)}(t)\,\tilde{K}(t)\,W(t)/m_{\mu}$
    as a function of Euclidean time $t$
    for the Y450 ensemble. In addition to the explanation of
    symbols in Figure~\ref{fig:integrand-G33}, we display the window LD$'$
    as defined in Equation~(\ref{eq:def:ld-prime-window}) (filled, orange downward triangles).}
    \label{fig:a33-disc-integrand-y450}
\end{figure}

A more quantitative statement can be obtained by isolating a contribution that
vanishes at leading and at $\mathcal{O}(\Delta m)$ by construction. In isoQCD the up and down
quark loops coincide configuration by configuration, so the disconnected
contribution to the isovector correlator, $G_{\mathrm{disc.}}^{(3,3)}$, in Equation~(\ref{eq:g33-disconnected-y450}) vanishes
identically in N451. The leading contribution to this term amounts to an
$\mathcal{O}(\Delta m^2)$ effect, and therefore it is  absent in the RM123 computation
at first order and is only accessible in the explicitly isospin-broken ensemble.
Its window decomposition on Y450 is shown in
Figure~\ref{fig:a33-disc-integrand-y450}, and the corresponding TMR integrals are
collected in Table~\ref{tab:results:hvp-33-38}. Rescaling to the
FLAG convention with the same factor $(6/14.026)^2$ as above, we obtain
$0.00031(3)$, $-0.00198(53)$ and $-0.33(62)$, in units of $10^{-10}$, for the
SD, ID and LD windows, respectively. In the short- and intermediate-distance
windows the contribution is resolved and lies two orders of magnitude below the
statistical uncertainty of the connected part, while in the long-distance window
it is compatible with zero with an uncertainty comparable to the bound in
Equation~(\ref{eq:a33-bound-dm2}).

One can make a sharper statement by defining a window in which the
$\mathcal{O}(\Delta m^2)$ contribution is most significant relative to the
uncertainty of the connected part. In Table~\ref{tab:results:hvp-33-38} we
additionally write the results in the ``LD$'$'' window, defined in
Equation~(\ref{eq:def:ld-prime-window}), where the disconnected contribution is
still resolved. In this window, the absolute value of the disconnected
contribution at the simulated up-down quark mass splitting amounts to about
$9\%$ of the statistical uncertainty of the connected part of N451. Rescaling
to the FLAG convention as above, this would amount to only $1.5\%$. Taken
together, this provides a direct determination of one higher-order contribution
and shows that higher-order effects in $\Delta m$ lie well below the current
level of precision.

For completeness, we also analyze the correlators using the bounding
method~\cite{RBC:2018dos}, along the lines of Reference~\cite{Ce:2022eix}. For
our particular ensembles, we use a single $\rho$ meson state as the upper bound
for both $G_{R}^{(3,3)}$ and $G_{R}^{(8,8)}$, and the results are displayed in
Table~\ref{tab:results:hvp-bounding}. While the LD$'$ window does not benefit
from this method, one can see that the uncertainties of the long distance window
are significantly reduced with respect to the results in
Table~\ref{tab:results:hvp-33-38}, which are computed directly from the raw
correlators. Even after this error reduction, which amounts to an order of
magnitude in the case of the isoscalar $a_{\mu}^{8,8}$, the results from both
ensembles are completely compatible with each other.\footnote{As explained in
Footnote \ref{fn:autocorr-barely-resolved}, the similar uncertainties in the LD
window of $a_{\mu}^{8,8}$ in both ensembles are a coincidence arising from a big
relative uncertainty in the determination of $\tau_{\mathrm{int}}$.}

\begin{figure}[t]
    \centering
    \includegraphics[width=0.49\textwidth]{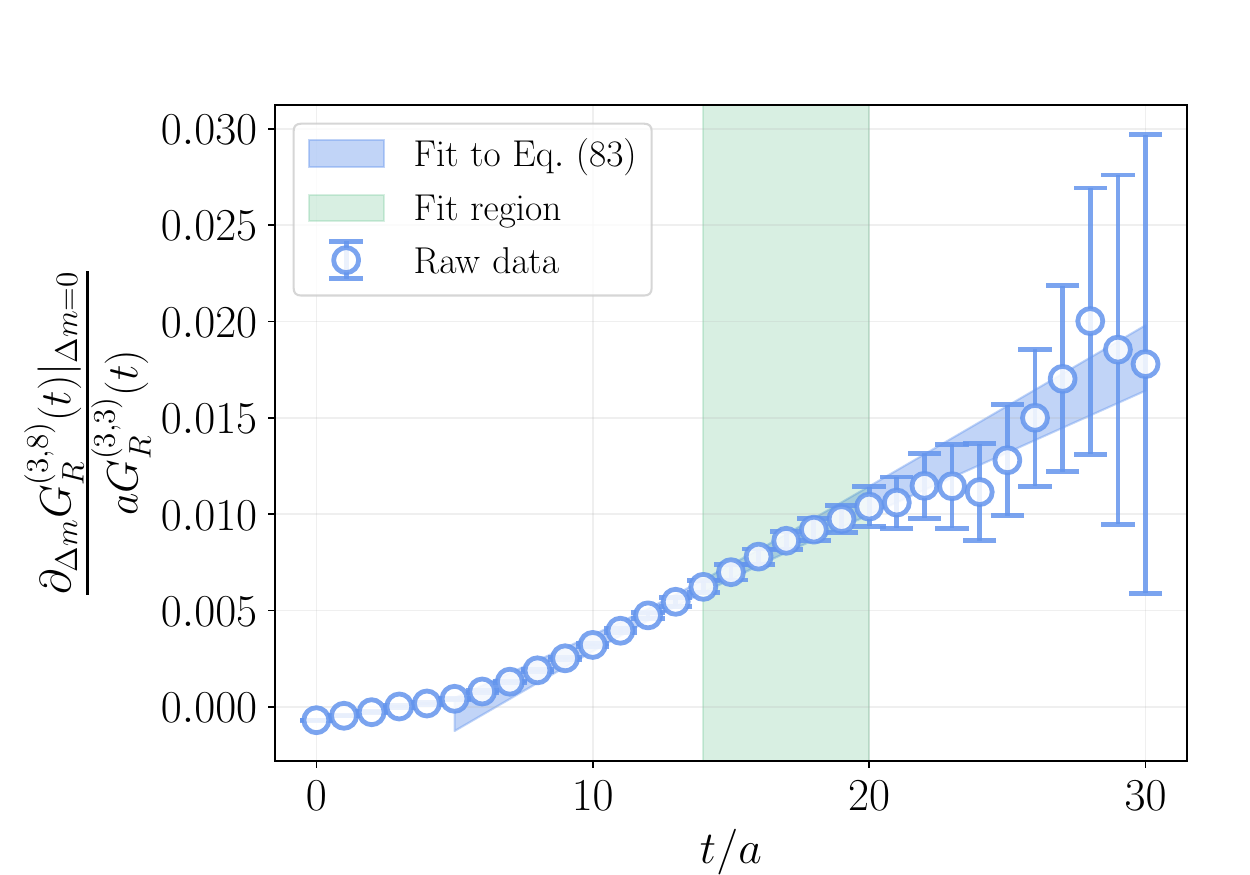}
    \caption{Ratio $\partial_{\Delta m} G_{R}^{(3,8)}(t)|_{\Delta m = 0} / G_{R}^{(3,3)}(t)$ (blue points) as a function of
      time in lattice units in the ensemble Y450, together with a fit to the
functional form in Equation~(\ref{eq:g38-spectral-form}) (blue band). The fit is
performed for the particular time region $t / a\in [14,19]$ (green band).}
    \label{fig:fit-c1c0}
\end{figure}

\begin{figure*}[t]
  \begin{subfigure}[t]{0.49\textwidth}
    \centering
    \includegraphics[width=\textwidth]{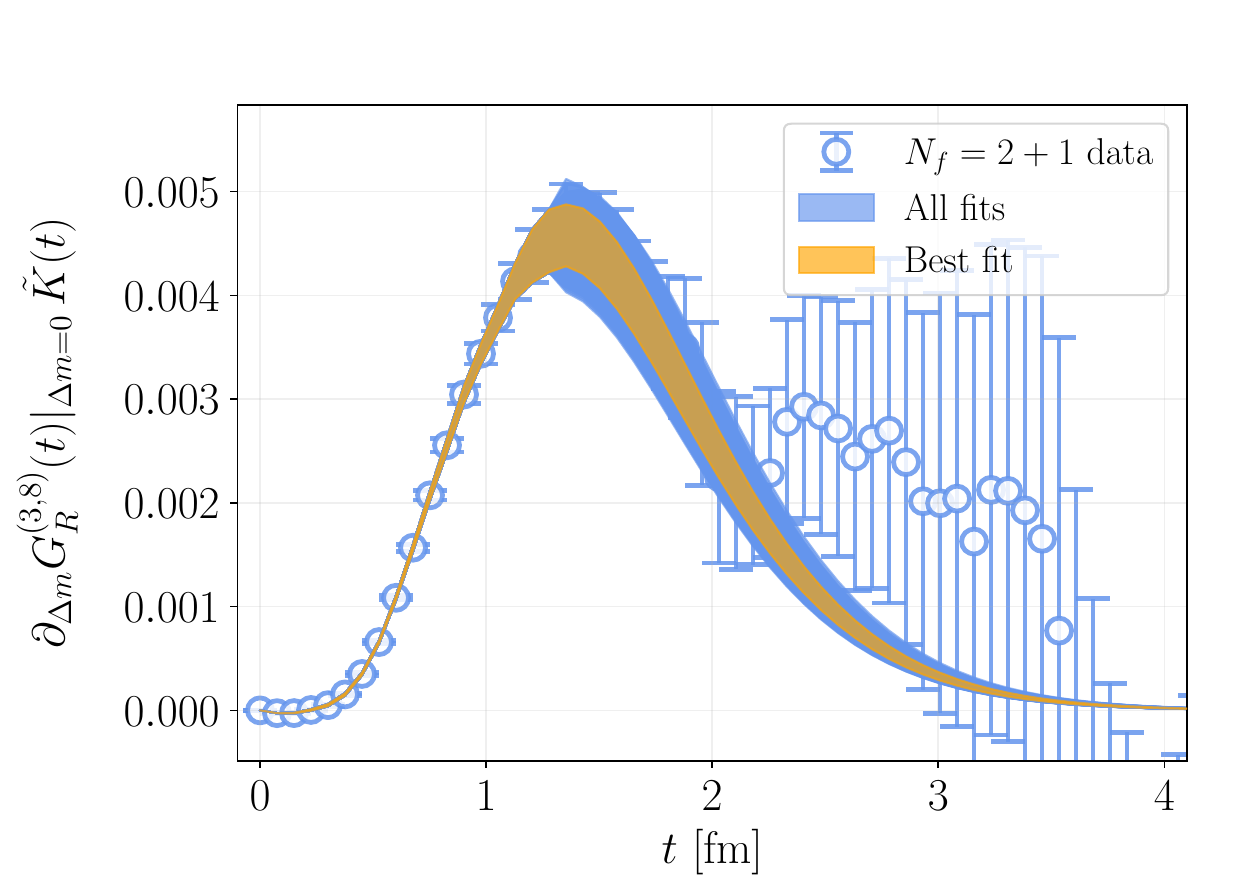}
  \end{subfigure}
  \hfill
  \begin{subfigure}[t]{0.49\textwidth}
    \centering
    \includegraphics[width=\textwidth]{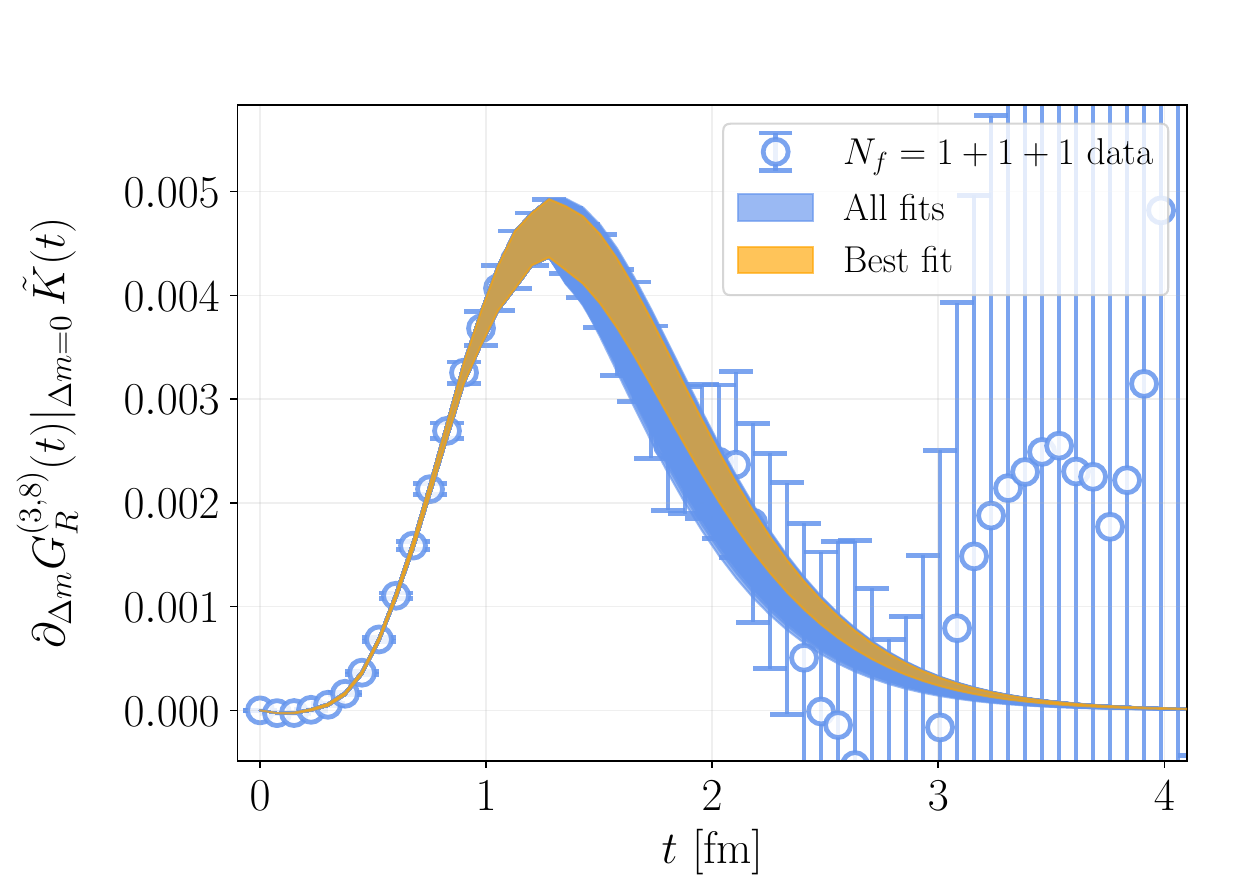}
  \end{subfigure}
  \caption{Integrand $\partial_{\Delta m} G_{R}^{(3,8)}(t)|_{\Delta m = 0}\,\tilde{K}(t)$,
for N451 (left) and Y450 (right). Blue points
correspond to raw lattice data, the blue band corresponds to different fits
coming from different choices of ranges in $t$, and the orange band corresponds
to the best fit.}
  \label{fig:integrand-G38}
\end{figure*}

\subsection{Isospin-violating contribution to the HVP and $R_{38K}$}

As we saw in Section~\ref{sec:correlator-construction}, the isospin-violating
contribution to the HVP contains disconnected diagrams which introduce a severe
signal-to-noise problem in the long-distance regime of the TMR integral. In contrast
to the case of $a_{\mu}^{8,8}$, it is not possible to obtain a result with significant
signal for $a_{\mu}^{3,8}$, and a method to describe the long-distance
tail of the correlator, based on its spectral reconstruction, is needed.

Since we know the spectral decomposition of the correlator from
Equation~(\ref{eq:g38-decomposition}), one can obtain $A$ and $B$ by fitting to the
functional form\footnote{The functional form in Equation~(\ref{eq:g38-spectral-form})
assumes open boundary conditions. For periodic boundary conditions, the ratio at
large times would take the form
\begin{align*}
A' + B' \left( t - \frac{T}{2} \right) \tanh \left[ m_{X} \left( t - \frac{T}{2} \right) \right],
\end{align*}
with $m_{X}$ being the mass of the lowest state in the energy tower, in this
case the $\rho$ meson. Since our ensembles have periodic boundary conditions,
this is the functional form used in the fits of this study, even though we keep
the slim notation of open boundary conditions in the text.}
\begin{align}
  \label{eq:g38-spectral-form}
\frac{\partial_{\Delta m} G_{R}^{(3,8)}(t; \Delta m)|_{\Delta m = 0}}{G_{R}^{(3,3)}(t)} \xrightarrow{t \to\infty} \bar{A} + \bar{B}t.
\end{align}
In Figure~\ref{fig:fit-c1c0} we show this ratio as a function of $t / a$, and observe
that the linear dependence is already visible at $t / a \sim 7$. In the figure,
we additionally show a fit to this linear form for a particular range in $t$,
from which we obtain the fit parameters $\bar{A}_{\mathrm{fit}}$ and
$\bar{B}_{\mathrm{fit}}$.

In order to obtain $R_{38K}$ with improved precision, we define the
reconstructed isovector-isoscalar correlator as
\begin{align}
  \partial_{\Delta m} G^{(3,8)}_{R,\text{rec.}} =
\begin{cases}
  \partial_{\Delta m}G_{R}^{(3,8)}|_{\Delta m = 0}, & t < t_{\text{cut}} \\
  (\bar{A}_{\text{fit}} + \bar{B}_{\text{fit}}t) G_{R}^{(3,3)}, & t \geq t_{\text{cut}}
\end{cases},
\end{align}
where $t_{\mathrm{cut}}$ defines the point above which the fitted functional form
is used instead of the raw lattice data. In practice, we found that the final
result for $R_{38K}$ is mostly sensitive to the fit parameter
$\bar{B}_{\mathrm{fit}}$. We account for possible systematic errors by performing
all possible fits between $t / a \in [10,25]$ with at least 5 data points, and
taking the half-width of the distribution of results of $R_{38K}$ obtained from
all fits.

In Figure~\ref{fig:integrand-G38} we show the integrand of $a_{\mu}^{3,8}$ with the
raw lattice data for $G^{(3,8)}$ as a function of $t$ for both $N_{f}=2+1$
(left) and $N_{f}=1+1+1$ (right) ensembles, and one can see that the signal is
lost at around $t \sim 2.5$ fm. As a blue band, we show the
integrands from all the fits coming from the different ranges in $t$ as
described above, together with an orange band representing the best fit. For all
fits and both ensembles, we chose $t_{\mathrm{cut}} = 1.3$ fm, and performing
the summation over $t$ and using Equation~(\ref{eq:r38k-definition}) we obtain
\begin{align}
  R_{38K} =& -2.16(15)(09)[17] \times 10^{-11} \text{ MeV}^{-1} \quad (\text{Y450}), \\
  R_{38K} =& -2.21(14)(12)[19] \times 10^{-11} \text{ MeV}^{-1} \quad (\text{N451}),
\end{align}
where the first and second errors are statistical and systematic, respectively.

One can see that the final errors for both ensembles are very similar,
even though N451 has double the number of configurations of
Y450. However, the systematic contribution to the error budget is
sizable and we do not consider it a clear indication for the
$N_{f}=1+1+1$ ensemble to be computationally advantageous. We also
note that the method we have used to extend the $G^{(3,8)}$ correlator
needs to be generalized at the physical pion mass, where one expects
the ratio (\ref{eq:g38-spectral-form}) to have a more complicated
functional form due to the $\pi\pi$ states below the $\rho$ mass.

Finally, following the FLAG convention introduced in Equation~(\ref{eq:flag-scheme}),
these results correspond to
\begin{align}
  \label{eq:amu-sib-flag}
  a_{\mu,\mathrm{SIB}}^{3,8} =&\; 12.96(90)(54)[1.02] \times 10^{-11} \quad (\text{Y450}), \\
  a_{\mu,\mathrm{SIB}}^{3,8} =&\; 13.26(84)(72)[1.14] \times 10^{-11} \quad (\text{N451}).
\end{align}
Once the charge factor of Equation~(\ref{eq:gamma-gamma-decomposition}) is included,
these values are of the same size as the phenomenological estimates of the
strong isospin-breaking contribution to the
HVP of References~\cite{James:2021sor,Hoferichter:2023sli}. We stress, however, that
neither agreement nor disagreement is to be expected, since our determinations
are obtained at a single lattice spacing, in a single volume and at an
unphysically heavy pion mass.

\section{Conclusions}
\label{sec:conclusions}


We have performed a direct comparison of two strategies to include strong
isospin-breaking effects in lattice QCD computations of the HVP: an
explicit $N_f = 1+1+1$ simulation and the Taylor expansion on
an $N_{f}=2+1$ simulation. Using two CLS ensembles only differing in the
splitting between up and down quark masses, we computed the kaon mass
difference, as well as the isovector ($a_\mu^{3,3}$), isoscalar
($a_{\mu}^{8,8}$), and isospin-violating ($a_{\mu}^{3,8}$) contributions to the
HVP of the muon $g-2$, where all correlators entering this comparison are renormalized
and fully $\mathcal{O}(a)$ improved. We have computed  $R_{38K}$
to better than $10\%$ precision on these ensembles,
a renormalized quantity that expresses the response of $a_\mu^{\rm hvp}$ to a kaon-mass splitting
induced by strong isospin breaking.

Our main findings are two-fold. On the one hand, we have found that, even with an
up-down quark mass splitting twice as large as the physical one, the 
expansion to first order in $\Delta m$ is compatible with explicit isospin broken
simulations, indicating that, 
at first order in $\Delta m$, it is robust even for high-precision physical determinations
such as the HVP contribution to $g-2$. On the other hand, the dynamical
simulation of strong isospin effects does not seem to lead to reduced
uncertainties in any of the observables studied in this work, in contrast to what
was found in Reference~\cite{RC:2025zoa} for the case of electromagnetic corrections.

Electromagnetic effects include valence, sea, and mixed
sea-valence effects at order $\mathcal{O}(e^2)$, and the study in
Reference~\cite{RC:2025zoa} indicates that the main contribution to the uncertainty
in the RM123 expansion approach comes from pure sea-quark contributions. Since the pure
sea-quark contributions are simulated in a dynamical QCD+QED setup (and are
therefore absent from the measurement of any observable), this would explain why
the explicit approach leads to smaller uncertainties in their case.  
For strong isospin breaking on the other hand, there are only valence effects at
$\mathcal{O}(\Delta m)$, valence-quark effects and pure sea-quark effects at
$\mathcal{O}(\Delta m^2)$, and mixed ones starting at
$\mathcal{O}(\Delta m^3)$ --- the latter would enter, for example, in the calculation
of the kaon-mass splitting beyond the linear order. Since the current level of
statistics is not sensitive to order $\Delta m^2$ or higher, this would provide
an explanation why a similar reduction in uncertainty has not been found in this
study. Nonetheless, from the difference in the results between the isoQCD and
the isospin-broken ensembles, one can estimate the higher-order effects in
$\Delta m$ to be $\lesssim 1 \times 10^{-10}$ for the isovector contribution
$a_{\mu}^{3,3}$. This is corroborated by a direct determination of the
disconnected contribution to $a_\mu^{3,3}$, which vanishes identically in isoQCD
and is purely $\mathcal{O}(\Delta m^2)$: in the long-distance window
between $1.0$ and $1.5$ fm we resolve it at the level
of $-0.075(12)\times 10^{-10}$ at the simulated kaon mass splitting, which, once
rescaled to the FLAG kaon-mass-splitting convention, amounts to $1.5\%$ of the statistical
uncertainty of $a_\mu^{3,3}$.

Taken together, our findings have a practical implication for future precision
computations of isospin-breaking corrections to the HVP: at the current level of
statistical precision, and for the observables considered here, investing
computational resources in $N_f = 2+1$ ensembles combined with the RM123
expansion appears to be a more efficient strategy than generating
dedicated $N_f = 1+1+1$ ensembles. This conclusion rests on the fact that
strong isospin breaking enters only through valence effects at first order and
on the smallness of the $\mathcal{O}(\Delta m^2)$ effects. It
may well change for observables where higher orders in $\Delta m$ are more
prominent, such as the pion-mass splitting and the low-energy
constant $\ell_{7}$~\cite{Frezzotti:2021ahg,Bonanno:2025cmh}, or where sea-quark
disconnected contributions play a larger role, both of which will be the subject
of future work. A systematic study of the $\mathcal{O}(\Delta m^2)$
and $\mathcal{O}(e^4)$ contributions using automatic differentiation
techniques~\cite{Albandea:2026cqh} is also planned.

\begin{table*}[t]
\centering
\begin{tabular}{lcc}
\toprule
 & N451 ($N_{f} = 2+1$) & Y450 ($N_{f} = 1+1+1$) \\
\midrule
\texttt{openQCD} version      & 1.6 & 2.4.2 \\
Light sector                  & TM Hasenbusch & RHMC \\
Strange sector                & RHMC & RHMC \\
Rational degree (light)       & --- & 16 \\
Rational range (light)        & --- & $[4\times10^{-4}, 8.15]$ \\
Rational degree (strange)     & 13 & 13 \\
Rational range (strange)      & $[5\times10^{-3}, 8.15]$ & $[5\times10^{-3}, 8.15]$ \\
Pseudofermions (light)        & 5 & $2\times 11$ \\
Pseudofermions (strange)      & 6 & 6 \\
Pseudofermions (total)        & 11 & 28 \\
$\tau$ [MDU]                  & 2.0 & 2.0 \\
Integrator levels             & 3 & 2 \\
Outer integrator              & OMF2, $8$ steps & OMF4, $14$ steps \\
\texttt{nstep} per level (inner $\to$ outer) & $1,\,1,\,8$ & $1,\,14$ \\
MDU between configurations    & 4 & 4 \\
$N_{\mathrm{cfg}}$            & 1011 & 450 \\
Acceptance                   & 94\% & 97\% \\
\bottomrule
\end{tabular}
\caption{Algorithmic parameters of the two CLS ensembles used in this work.
  The light sector of N451 is simulated with twisted-mass Hasenbusch frequency
  splitting, while in Y450 each of the three flavors is simulated with its own
  single-flavor RHMC factorization. The rational range is the spectral interval
  $[r_{a}, r_{b}]$ of $|\gamma_{5}\hat{D}|$ of the Zolotarev
  approximation.}
\label{tab:rhmc-parms}
\end{table*}

\begin{acknowledgments}


The authors would like to thank Konstantin Ottnad for providing the data of the
disconnected quark loops of the N451 ensemble and providing the code to compute
them in the Y450 ensemble. We thank Andreas Risch for providing the code to
compute the connected diagrams of this work and for useful discussions. We thank
Dalibor Djukanovic for useful discussions during the evolution of the project
and during the writeup of the paper. We also thank Alessandro Conigli,
Alessandro Barone and Dominik Erb for useful discussions. We are grateful to our
colleagues in the CLS initiative for sharing ensembles.

The authors gratefully acknowledge the computing time made available to them on
the high-performance computer Mogon-NHR at the NHR Center NHR S\"ud-West. This
center is jointly supported by the Federal Ministry of Research, Technology and
Space and the state governments participating in the National High-Performance
Computing (NHR) joint funding program (\url{https://go-nhr.de/our-partners}).
Additional calculations for this project have been performed on the HPC cluster
Mogon-II at Johannes Gutenberg-Universit\"at Mainz.

This work has been supported by Deutsche Forschungsgemeinschaft
(German Research Foundation, DFG) through the Collaborative Research
Center 1660 ``Hadrons and Nuclei as Discovery Tools'', and through the
Cluster of Excellence ``Precision Physics, Fundamental Interactions
and Structure of Matter'' (PRISMA++ EXC 2118/2), funded within the
German Excellence Strategy (Project No.\ 390831469), as well as through the Emmy
Noether Programme (Project No. 559290516).

Our programs use the \texttt{QDP++} library~\cite{Edwards:2004sx} and the deflated SAP+GCR
solver from the \texttt{openQCD} code~\cite{Luscher2019_openQCD}. The Wick contractions
displayed and computed in this work have been obtained with the \texttt{Quark
Contraction Tool}~\cite{Djukanovic:2016spv}.

\end{acknowledgments}


\appendix

\section{RHMC simulation details}
\label{app:rhmc}

\subsection{RHMC parameters}
\label{sec:rhmc-parameters}

Both ensembles were generated with the \texttt{openQCD}
code~\cite{Luscher:2012av,Luscher2019_openQCD}, and differ in the way the
light-quark determinant is factorized into pseudofermion actions. Here we
summarize the simulation parameters of both setups, which are collected in
Table~\ref{tab:rhmc-parms}.

For a single quark flavor, \texttt{openQCD} uses even-odd preconditioning and
starts from the decomposition~\cite{Luscher2019_openQCD}
\begin{align}
  \det D = \det(1_{e} + D_{oo}) \, \det \hat{D},
\end{align}
where $\hat{D}$ is the even-odd preconditioned Dirac operator and $D_{oo}$ its
odd-odd part. The first factor is included directly in the molecular-dynamics
Hamiltonian, while the second one is written as
\begin{align}
  \label{eq:rhmc-factorization}
  \det \hat{D} = W \det R^{-1}, \qquad W = \det(\hat{D} R),
\end{align}
with $R$ a Zolotarev rational approximation to $(\hat{D}^{\dagger}\hat{D})^{-1/2}$
of degree $[n,n]$,
\begin{align}
  R = r_{b}^{-1} R_{n,\epsilon}(r_{b}^{-2}\hat{D}^{\dagger}\hat{D}), \qquad
  \epsilon = (r_{a}/r_{b})^{2}.
\end{align}
The residual factor $W$ is treated by an additional reweighting factor.

In the $N_{f} = 1+1+1$ ensemble Y450, the three flavors are simulated as three
independent single-flavor RHMC factorizations. For the up and the down quark a
rational approximation of degree $n = 16$ in the range $[4\times 10^{-4},
8.15]$ is used. Each light flavor is split into one factor containing the six
poles with the largest shifts, which also carries the $\det(1_{e}+D_{oo})$ term,
and ten factors containing a single pole each, amounting to $11$ pseudofermion
fields per light flavor. The strange quark is split into one factor with eight
poles and five single-pole factors, i.e.\ $6$ pseudofermion fields, for a total
of $28$.

In the $N_{f} = 2+1$ ensemble N451 the degenerate light quarks form a pair
with positive determinant and can therefore be simulated with twisted-mass
Hasenbusch frequency splitting~\cite{Hasenbusch:2001ne,Luscher:2008tw},
distributed over five pseudofermion fields. The strange quark is treated with
exactly the same RHMC factorization as in Y450, with $6$ pseudofermion fields,
giving $11$ in total.  The explicit splitting of the light sector therefore
replaces one comparatively cheap twisted-mass doublet by two single-flavor RHMC
chains and increases the number of pseudofermion fields by a factor
of $28/11 \approx 2.5$.

Both ensembles use a trajectory length $\tau = 2$ in molecular-dynamics units
(MDU) and a hierarchical integrator, in which the forces assigned to the
innermost level are integrated with the finest step. In N451 the lightest
twisted-mass factor is placed at the outermost level, integrated with $8$ steps
of a second-order Omelyan-Mryglod-Folk (OMF2) integrator ($\lambda = 1/6$),
leading to an acceptance of 94\%. In Y450 all $28$ fermion forces are integrated
at a single level with $14$ OMF4 steps, with the gauge force at the inner level,
leading to an acceptance of 97\%. Taken together, we saw that in practice the
cost of generating one configuration of Y450 is approximately three times the
cost of one configuration of N451.

\subsection{Reweighting the negative sign of the determinant of the Wilson-Dirac operator}
\label{sec:reweighting}

\begin{figure}[t]
    \centering
    \includegraphics[width=0.49\textwidth]{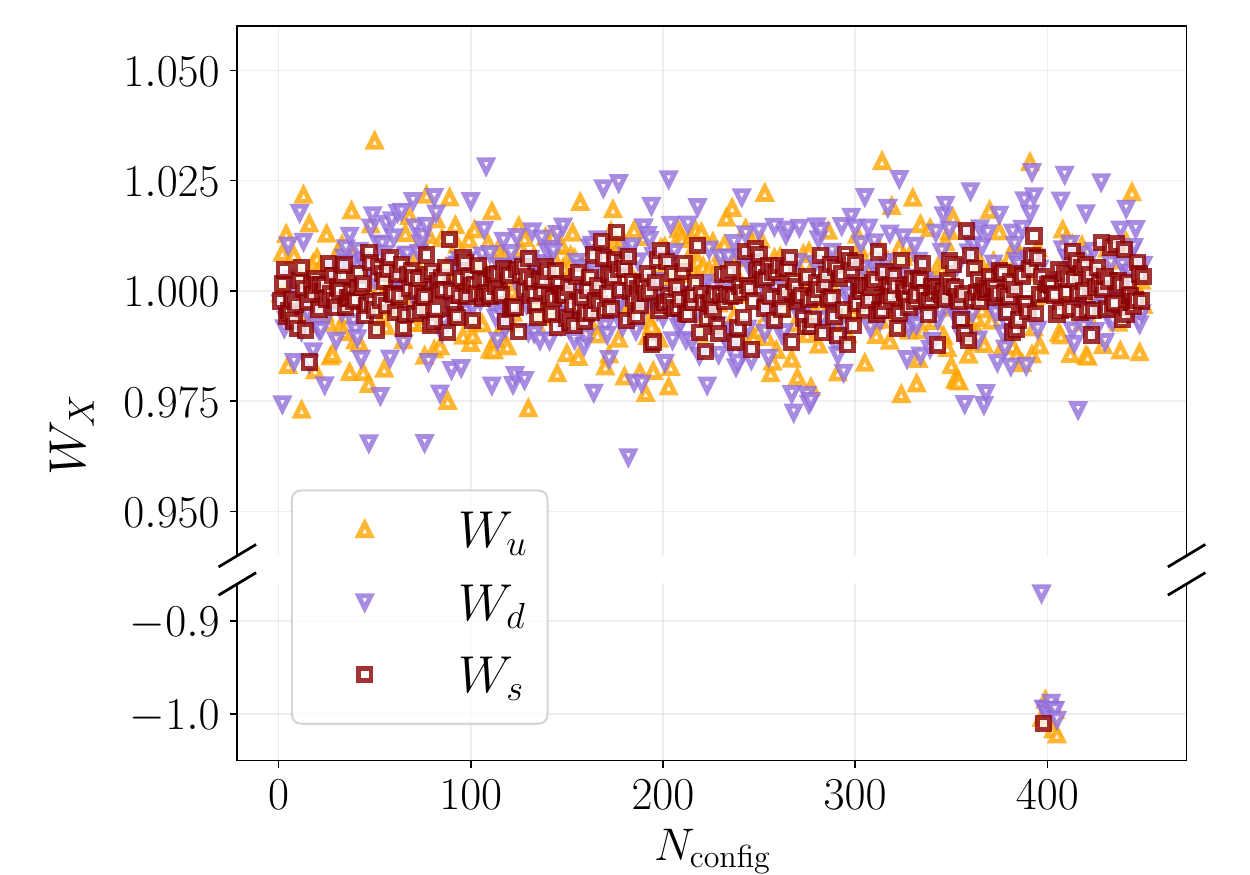}
    \caption{Reweighting factors $W_{u}$, $W_{d}$ and $W_{s}$ of the three
      single-flavor RHMC factorizations of the $N_{f} = 1+1+1$ ensemble Y450,
      as a function of the configuration index $N_{\mathrm{config}}$, with the
      two replica shown consecutively. The upper
      frame collects the configurations with $\det Q_{X} > 0$ and the lower one
      those with $\det Q_{X} < 0$. }
    \label{fig:reweighting-factors}
\end{figure}

Let us consider the path integral of a single fermion flavor $\psi$,
\begin{align}
\langle O \rangle =&\; \frac{1}{Z} \int \mathcal{D}U \mathcal{D}\psi \mathcal{D}\bar{\psi}  \; O[U, \psi, \bar{\psi}] \det D
                     \;e^{-S_{G}} e^{-\bar{\psi} D \psi}  \nonumber \\
  =&\; \frac{1}{Z} \int \mathcal{D}U \; O[U] \det Q \;e^{-S_{G}}
\end{align}
where $Q = \gamma_{5}D$, and we have used $\det \gamma_{5} = 1$ and the
identity\footnote{Note that this identity is always true regardless of $D$
having negative eigenvalues
(or $D^{-1}$ existing at all)~\cite{Gattringer:2010zz}.}
\begin{align}
\det D = \int \mathcal{D}\psi \mathcal{D}\bar{\psi} \; e^{-\bar{\psi} D \psi}.
\end{align}
The Zolotarev approximation in the RHMC algorithm does
not approximate $\det Q$, but $\det \left| Q \right|$, where
 $\left|  Q \right| \equiv \sqrt{Q^{\dagger}Q}$ is a positive semi-definite
matrix. Since $Q$ is Hermitian and can be decomposed
as $Q = V \Lambda V^{\dagger}$, with $V$ unitary and $\Lambda$ diagonal, it also
follows that $\left| Q \right| = V \left| \Lambda \right| V^{\dagger}$.
Therefore, the integral in the pseudofermion representation,
\begin{align}
\det \left| Q \right| = \int \mathcal{D}\phi\; e^{-\phi^{\dagger} \frac{1}{\sqrt{Q^{\dagger}Q}} \phi},
\end{align}
where now $\phi$ is a bosonic field, is convergent for any gauge field $U$. The
function $f(Q^{\dagger}Q) = 1 / \sqrt{Q^{\dagger}Q}$ can be built by means of a
Zolotarev approximation~\cite{Luscher2019_openQCD}.

We can therefore rewrite $\langle O \rangle$ as
\begin{align}
  \label{eq:reweight-sign-detD-path-integral}
\langle O \rangle = \frac{\int \mathcal{D}U\mathcal{D}\phi\; O[U] \frac{\det Q}{\left| \det Q
\right|} e^{-S_{G} - \phi^{\dagger} \frac{1}{\sqrt{Q^{\dagger}Q}}\phi}}{\int \mathcal{D}U\mathcal{D}\phi\;  \frac{\det Q}{\left| \det Q
\right|} e^{-S_{G} - \phi^{\dagger} \frac{1}{\sqrt{Q^{\dagger}Q}}\phi}}
\end{align}
where we used that
$\det \left| Q \right| = \det \left|  \Lambda \right| = \left|  \det \Lambda \right| = \left|  \det Q \right|$.
From Equation~(\ref{eq:reweight-sign-detD-path-integral}), one can clearly see that
gauge configurations can be generated
following $e^{-S_{G}}\left| \det Q \right|$, while
the sign of the Dirac
operator $\mathrm{sign}(\det Q) = \det Q / \left|  \det Q \right|$ is taken into
account as a reweighting factor.

The sign of $\det Q$ was determined configuration by configuration from the
low-lying spectrum of the even-odd preconditioned Wilson–Dirac
operator $\hat D$, along the lines of Reference~\cite{Kuberski:2023zky}.
Particularly, it was computed with a version of \texttt{openQCD} 2.0 interfaced
to the \texttt{SLEPc}~\cite{Hernandez:2005slepc,Roman:2020slepc} eigensolver
library, which relies on
\texttt{PETSc}~\cite{Balay:2020petsc,Balay:1997efficient}.

The resulting reweighting factors for the three
flavors of Y450 are shown in Figure~\ref{fig:reweighting-factors}. Note that nine
configurations, consecutive in Monte Carlo time, carry $\det Q_{u} < 0$
and $\det Q_{d} < 0$ simultaneously. However, only one of them also
has $\det Q_{s} < 0$, and therefore the product of the
three leads to a negative product of determinants in only one configuration.

Finally, Figure~\ref{fig:reweighting-factors} shows one configuration with
$W_{d} \approx 0.87$. Removing configurations one at a time from the analysis of
the correlators singles out this same configuration, and excluding it reduces
their statistical errors significantly. Following the discussion in
Reference~\cite{Agadjanov:2023efe}, we exclude it from the measurements,
treating it as a missing configuration in the $\Gamma$ method.

\newpage
\bibliography{biblio}

\end{document}